# An Enhanced Multi-Objective Biogeography-Based Optimization for Overlapping Community Detection in Social Networks with Node Attributes

Ali Reihanian[a,c], Mohammad-Reza Feizi-Derakhshi[b], Hadi S. Aghdasi[b]

[a] *Department of Computer Engineering, Faculty of Engineering, Arak University, Arak, 38156-8-8349, Iran*
[b] *Department of Computer Engineering, Faculty of Electrical and Computer Engineering, University of Tabriz, Tabriz, Iran*
[c] *Research Institute for Artificial Intelligence, Arak University, Arak, 38156-8-8349, Iran*

## Abstract

Community detection is one of the most important and interesting issues in social network analysis. Most of the current community detection algorithms tend to find communities in social networks with just considering the topological structures of the networks. In recent years, simultaneously considering of nodes' attributes and topological structures of social networks in the process of community detection has attracted the attentions of many scholars, and this consideration has been recently used in some community detection methods to increase their efficiencies and to enhance their performances in finding meaningful and relevant communities. But the problem is that most of these methods tend to find non-overlapping communities, while many real-world networks include communities that often overlap to some extent. In order to solve this problem, an evolutionary algorithm called MOBBO-OCD, which is based on multi-objective biogeography-based optimization (BBO), is proposed in this paper to automatically find overlapping communities in a social network with node attributes with synchronously considering the density of connections and the similarity of nodes' attributes in the network. In MOBBO-OCD, an extended locus-based adjacency representation called OLAR is introduced to encode and decode overlapping communities. Based on OLAR, a rank-based migration operator along with a novel two-phase mutation strategy and a new double-point crossover are used in the evolution process of MOBBO-OCD to effectively lead the population into the evolution path. In order to assess the performance of MOBBO-OCD, a new metric called *alpha_SAEM* is proposed in this paper, which is able to evaluate the goodness of both overlapping and non-overlapping partitions with considering the two aspects of node attributes and linkage structure. Quantitative evaluations, based on three extensive experiments on 14 real-life data sets with diverse characteristics, reveal that MOBBO-OCD achieves favorable results which are quite superior to the results of 15 relevant community detection algorithms in the literature.

## 1. Introduction

Almost any natural phenomena can be modeled as networks by defining a set of entities and establishing a criterion of the relation between them [32]. A social network can be considered as a well-known example of a network, which is a social structure made up of a set of nodes as the social actors who are connected by one or more specific types of interdependency [3]. Community is a significant substructure in many complex networks [27]. Since social networks are considered as a kind of complex networks, their community structure is one of their distinctive properties, which can reveal their organization and the hidden relation among their components [16]. Identifying meaningful communities of social networks is an interesting field of study which has attracted many researchers in recent years [31]. A community can be defined as a subset of nodes that are densely connected to each other and loosely connected to the nodes in the other communities in the same network [19], like a group of individuals in a social network who are friends with each other. Since it is more likely that the members of a community have common hobbies, social functions, etc., the identified communities can be used in collaborative recommendation, information spreading, knowledge sharing, and other applications that are beneficent for us [47].

Community detection, also known as graph clustering, is one of long-standing popular research topics [13]. Most of the researches in the field of community detection focused on designing a variety of methods for non-overlapping (disjoint or separated) community detection, in which every node just belongs to exactly one community [46]. However, many real-world networks include communities that often overlap to some extent. It means that some nodes of these networks may belong to more than one community because they may have different roles in the network [46]. For example, we can consider an individual in a social network that might be a member of a karate community and a cinema community, simultaneously.

On the other hand, most of the studies in the field of community detection focused on the graph structures of social networks to detect communities, while no content analysis is performed in their process of community detection [31]. In many real-world social networks, there is one or more attributes assigned to each node, which describe its properties, and are often homogenous in a community [18]. In other words, it is more likely that the nodes with the same attributes belong to the same communities. Nowadays, real world networks contain a vast range of information which can be classified as node (user) attributes, such as shared objects, comments, following information, age, education, gender, profession, etc [29]. Thus, the process of community detection can be more optimized with considering contents of a social network (if available) in finding communities in which members are not just densely connected but share similar attributes [18].

The problem of overlapping community detection have been considered in some researches, and some efficient overlapping community detection methods have been proposed in the literature of community detection in which no content analysis are performed. On the other hand, in recent years, the interest of scholars for finding community structures of social networks with considering node attributes and link structure have increased which have led them to propose some non-overlapping community detection methods. But, to our best knowledge, the problem of detecting overlapping communities in social networks with node attributes with synchronously considering structure and attribute is remained as an open problem.

In order to solve the mentioned problem, in this paper, we proposed a multi-objective evolutionary algorithm called MOBBO-OCD to automatically find overlapping communities in a social network, in which node attributes are available, with synchronously considering the density of connections and the similarity of nodes' attributes. Our proposed algorithm is based on biogeography-based optimization (BBO) [35], which is a novel promising evolutionary algorithm proposed with inspiration from the science of biogeography to solve global optimization problems. Since attribute similarity and connection density can be considered as two independent and sometimes conflicting objectives [18], we used a multi-objective BBO to make balance between them. The final result of the proposed method is a set of non-dominated solutions (partitions of a network) which contain partitions which have the best performance from the perspective of topological structure (density of connections) of a network, partitions which have the best performance from the perspective of similarity of nodes' attributes in the network, and partitions that reach to a trade-off between the density of connections and the similarity of nodes' attributes in the network. Thus, our proposed method can provide a wide range of solutions for a decision maker to choose from. With considering the described goals and characteristics of MOBBO-OCD, The following contributions are made in this paper:

- Since MOBBO-OCD aims to detect overlapping communities, we present an extended locus-based adjacency representation called OLAR to encode and decode overlapping communities. OLAR performs in 4 stages, which are Encoding, First Decoding, Marking and Final Decoding.
- Based on OLAR, we employ a rank-based migration operator, introduce a two-phase mutation strategy, and propose a double-point crossover in the evolution process of MOBBO-OCD to effectively lead the population into the evolution path.

- We propose a performance metric called *alpha_SAEM*, which is able to evaluate the goodness of both overlapping and non-overlapping partitions with considering the two aspects of node attributes and linkage structure.
- We conduct three extensive experiments on 14 real-life data sets with different characteristics to evaluate the performance of MOBBO-OCD by comparing its results with those of 15 relevant community detection algorithms.

The remainder of this paper is outlined as follows: Section 2 reviews the related works. In Section 3, the background of the research is explained. Section 4 presents the proposed algorithm MOBBO-OCD, and explains it in detail. In order to evaluate the performance of MOBBO-OCD, extensive experiments are conducted on 14 real-life data sets. The descriptions of these data sets, the experimental results and their analysis are given in Section 5. Finally, Section 6 concludes the paper.

## 2. Related works

Many studies have been made in the area of community detection, in which the majority mainly focuses on the topological structures or linkage patterns of networks for finding their communities [47]. Considering the community detection strategies which are employed in these studies, their proposed methods can be classified into non-overlapping (separated or disjoint) community detection methods and overlapping community detection methods [30].

Palla et al. [25] proposed an approach called CPM to analyze the main statistical features of the interwoven sets of overlapping communities. CPM is performed based on first locating all cliques (maximal complete sub-graphs) of a network, and then identifying its communities by carrying out a standard component analysis of the clique-clique overlap matrix. Gregory [11] presented an algorithm called COPRA for finding overlapping community structures of very large networks, which is based on the label propagation technique introduced in [28] but is able to detect communities that overlap. Cazabet and Amblard [4] proposed a multi-agent system that aims to simulate both the evolution of a network and the joint evolution of communities on it. This system can deal with overlapping communities. Lancichinetti et al. [17] presented a method called OSLOM that is capable of detecting clusters in networks accounting for edge directions, edge weights, overlapping communities, hierarchies and community dynamics. OSLOM is based on the local optimization of a fitness function expressing the statistical significance of clusters with respect to random fluctuations. Xie et al. [39] introduced a dynamic interaction process to allow efficient and effective overlapping community detection. SLPA is a specific implementation of this dynamic process. Yang and Leskovec [41] proposed Community-Affiliation Graph Model, which is a model-based community detection method that builds on bipartite node-community affiliation networks. This method can successfully capture overlapping, non-overlapping as well as hierarchically nested communities. Also, Yang and Leskovec [42] presented an overlapping community detection method called BIGCLAM that scales to large networks of millions of nodes and edges. Yang et al. [44] introduced a method for overlapping community detection called CoDA that scales to networks with millions of nodes and tens of millions of edges. CoDA exhibits the following three properties: detecting both cohesively connected communities (in which nodes link to each other) as well as 2-mode communities (in which nodes link in a bipartite fashion, where links predominate between the two partitions rather than inside them), allowing cohesive and 2-mode communities to overlap or be hierarchically nested, and allowing for community detection in directed as well as undirected networks. Xing et al. [40] proposed a novel overlapping method, named as ONS-OCD, which takes the high quality disjoint community structure as the input. At the first step, the algorithm uses the node similarity to get the potential members of each community. At the second step, the potential members of every community are analyzed, and the influence of the nodes on the community is calculated. Finally, at the last step, the final overlapping nodes are achieved based on the node influence on communities. Jokar et al. [15] used the balanced link density label propagation (BLDLP) algorithm [14] and the fuzzy theory to propose a novel method for overlapping community detection, called fuzzy BLDLP. Their proposed method needs no prior information about the community numbers of a network in order to discover its communities. Doluca and Oguz [10] proposed a novel method called Adjacency Propagation Algorithm (APAL) which considers the notion that the adjacent vertices are the best candidates for detecting overlapping communities in an undirected, unweighted, nontrivial graph. APAL is a compact overlapping community detection algorithm with a single threshold parameter used to filter the detected communities according to their intra-connectivity property. Ding et al. [9] introduced a robust local-expansion-based overlapping community detection algorithm, named CEO, performing Construction, Expansion and Optimization sub-processes. To solve the poor fault tolerance problem, CEO discards low quality seeds and communities in each sub-process based on optimizing node memberships. Yang et al. [45] proposed an overlapping community detection method, called QOCE (Quadratic Optimization based Clique Expansion), that does not require any prior knowledge. QOCE follows the popular seed set expansion strategy and regards each high-quality maximal clique as the initial seed set. For seed set expansion, QOCE uses a fast short random walk to sample a sub-graph from a clique seed set, and then, adopts a quadratic optimization to approximate the Cheeger cut on the sampled sub-graph. Finally, a local minimum of conductance determines the boundary of the community. Since community detection can be viewed as a clustering optimization problem, evolutionary computation and swarm-intelligence-based algorithms have a chance to be used for solving the community detection problem [30]. Compared with traditional algorithms, intelligent optimization algorithms can effectively find a proper, high-quality solution within a reasonable period of time [5]. Tian et al. [37] introduced an evolutionary multi-objective optimization based fuzzy method, named as EMOFM, for overlapping community detection. The proposed method optimizes the community centers by using a specially tailored multi-objective evolutionary algorithm, and also, it can find an appropriate fuzzy threshold for each node, so that diverse overlapping community structures can be uncovered. Ma et al. [23] proposed a local-to-global scheme-based MOEA (multi-objective evolutionary algorithm), to detect overlapping communities of large-scale complex networks. The proposed method is called LG-MOEA, which includes a local community structure detection stage and a community structure determination stage. Shang et al. [33] proposed an evolutionary multi-objective algorithm based on similarity matrix and node correction for overlapping community detection. Their proposed algorithm can solve the high sensitivity of the community centers' initialization, and also, it can correct misclassified nodes.

The most important drawback of the studies which were described in the previous paragraph is that they merely consider the graph structure of networks for finding their communities. But, nowadays, real world networks are containing a vast range of contents, like node attributes, which can be analyzed and used to enhance the process of community detection. Thus, it is unreasonable for a community to be explained by a single entity [29].

In recent years, several studies have presented approaches which consider both the contents that are interchanged in networks and the topological structures of the networks, in order to find more meaningful communities [29]. Zhao et al. [47] proposed a topic-oriented community detection approach based on social objects' clustering and link analysis. Their introduced approach can identify the communities which reflect topics and strengths of connections, simultaneously. Reihanian et al. [31] evaluated the effect of topic consideration for finding meaningful communities in rating-based social networks. They presented a topic-oriented community detection algorithm based on the method proposed in [47] for detecting communities of rating-based social networks. With conducting experiments on real life data sets, they came to this conclusion that the results of community detection in rating-based social networks will be improved when the topic of interest is considered. Xia and Bu [38] constructed a semantic network from semantic information extracted from user-comment contents, and then implemented a community-detection algorithm on the giant component of the constructed semantic network in order to find communities. Yang et al. [43] presented an algorithm, named CESNA, for detecting overlapping communities in networks with node attributes. CESNA statistically models the interaction between the network structure and the node attributes. Papadopoulos et al. [26] presented an iterative parallelizable algorithm called CLAMP for clustering attributed multi-graphs, which automatically balances the structural and attribute properties of the vertices, and clusters the network such that objects in the same cluster are characterized by similar attributes and connections. Reihanian et al. [29] proposed a multi-objective discrete biogeography-based optimization algorithm to find non-overlapping communities of social networks with node attributes. Their method tends to reach to a trade-off between similarity of nodes' attributes and density of connections in the identified communities. Li et al. [18] introduced a multi-objective evolutionary algorithm called MOEA-SA, which is based on structural and attribute similarities, to solve the attributed graph clustering problem. Reihanian et al. [30] introduced a generic framework called SNTOCD to detect overlapping communities in social networks, with special focus on rating-based social networks. Their proposed framework considers the information shared by the users (ratings), as well as their topics of interest, for the sake of finding meaningful communities. Ma et al. [21] proposed a dual-population-based multi-objective evolutionary algorithm, called DP-MOEA, for balancing topology structure and node attribute in community detection of attributed networks. In DP-MOEA, one population takes charge of community detection according to the topology structure, whereas the other population is responsible for community detection based on the node attribute information. The two populations evolve independently by different genetic operations and interact with each other at every certain number of generations to utilize the good individuals obtained in the other population. Moreover, a node attribute similarity-based local search strategy and a community merging strategy are designed in the procedure of population interaction to enable the generation of high-quality individuals. Chen et al. [6] proposed a new community detection approach in attributed networks, called SOA, by examining the Subspaces of Attributes. Their basic idea was to regard the community detection as an optimization problem, which aimed at learning a new similarity matrix to maximize the homophily. Zhao et al. [48] proposed a community detection method, which not only considers both structural topology and node content, but also provides a flexible parameter to balance their contribution. Zhou et al. [50] proposed a community detection algorithm, called CDBNE, based on unsupervised attributed network embedding. Actually CDBNE is a framework that learns the representation based on network structure and attribute information and the clustering-oriented representation, simultaneously. The framework includes the graph attention auto-encoder module, the modularity maximization module and the self-training clustering module.

Some of the methods described in this section synchronously consider structure and attribute in finding communities of social networks with node attributes. But most of these methods, such as the methods described in [18, 29], can find disjoint communities, and are not able to find overlapping ones, while the other methods, like the one introduced in [26], need to know the number of communities beforehand, while the number of communities in a real graph is usually unknown in advance [18]. In order to overcome these problems, we propose a novel community detection algorithm in this paper, which can automatically find overlapping communities with synchronously considering the density of connections and the similarity of nodes' attributes in a social network with node attributes.

## 3. Background

In this section, Biogeography-Based Optimization (BBO) and Multi-objective Optimization Problem (MOP), as the concepts which organize the background of this research, are discussed.

### 3.1. Biogeography-Based Optimization (BBO)

As previously mentioned, biogeography-based optimization (BBO) is an evolutionary algorithm, which was introduced by Dan Simon in 2008 [35], to solve global optimization problems. BBO is based on the science of biogeography, which deals with the study of the distribution of biological species over time and space [20, 29, 35].

In BBO, each member of the population is called the habitat, which represents a candidate solution to the problem that tends to be solved by BBO. Each habitat includes a *n*-dimensional vector, and each of the *n* variables of this vector is a value of the so-called Suitability Index Variable (SIV) [29]. A measure of the goodness of a habitat in BBO is its Habitat Suitability Index (HSI) [20]. A good habitat has a high value of HSI. HSI is equivalent to fitness, in maximization problems, or cost, in minimization problems, in some other evolutionary algorithms [29]. Like many other evolutionary algorithms, there are two main steps in BBO which are [20]: 1) information sharing and 2) mutation.

Migration operator is introduced in BBO to perform the step of information sharing, which tends to improve the habitats of the population. Sharing SIV values between habitats is a probabilistic task in BBO which is conducted based on the migration rates of the habitats. The migration rates of a habitat like $H_i$ have two components: immigration rate ( $\lambda_i$ ) and emigration rate ( $\mu_i$ ). The immigration rate of a habitat is used to probabilistically decide whether to immigrate or not (that means whether to accept an SIV value from other habitats or not). If the immigration is selected for the habitat $H_i$, then the emigrating habitat should be selected (a habitat which its SIV value should be transferred). The emigrating habitat $H_j$ is probabilistically selected based on its emigration rate ( $\mu_j$ ). In BBO, each call of the migration operator can lead to migration of a single SIV value from one habitat (emigration) to another habitat (immigration) of the population [20, 29].

The mutation in BBO is a probabilistic operator which randomly changes an SIV value of a habitat. The aim of the mutation is to increase diversity among the population (habitats). Mutation gives this chance to the habitats with low HSIs to improve their quality (enhance their HSI values). On the other hand, the habitats with high HSIs are also given this chance by mutation to improve themselves even more than they already have [20].

Simon et al. [36] showed that biogeography-based optimization (BBO) is a generalization of a genetic algorithm with global uniform recombination (GA/GUR), but they also mentioned that one BBO characteristic which makes it distinctive from GA/GUR is its migration mechanism, which affects selection pressure (i.e., the probability of retaining certain features in the population from one generation to the next). They also demonstrated that the unique selection pressure provided by BBO generally results in better optimization results for a set of standard benchmark problems. Thus, we can conclude that GA is a general framework, and BBO explains particular and interesting specifications of the GA framework.

In this paper, we proposed a multi-objective evolutionary algorithm based on BBO, which is called MOBBO-OCD, to automatically find overlapping communities in a social network, in which node attributes are available, with synchronously considering the density of connections and the similarity of nodes' attributes. The motivation of utilizing BBO as the optimization algorithm in the proposed method (MOBBO-OCD) is that BBO is one of the fastest-growing nature-inspired algorithms for solving practical optimization problems, and has the advantages in terms of simplicity, flexibility and computational efficiency [22]. Since its introduction, BBO has been employed in different researches to solve numerous practical optimization problems in various branches of science and engineering, such as data analysis, network and antenna problems and image processing [22].

### 3.2. Multi-objective Optimization Problem (MOP)

In this sub-section, first, a formal definition of an MOP is given. Then, the concept of dominance is explained. After that, the Pareto-based approach (one of the most considered general approaches for solving an MOP), which is employed in this paper, is discussed.

#### 3.2.1. A formal definition of an MOP

An MOP can be formally defined as: “minimizing (or maximizing) $F(x) = \{f_1(x),..., f_k(x)\}$. An MOP solution minimizes (or maximizes) the components of a vector $F(x)$, where *x* is a *n*-dimensional decision variable vector $x = \{x_1,..., x_n\}$ from some universe $\Omega$ [7].”

#### 3.2.2. The concept of dominance

In a multi-objective minimization problem, solution *x* is said to dominate solution *y* ( $x \prec y$ ), if:

$$\forall i: \mathrm{f_i(x)} \le f_i(y) \qquad (1)$$

$$\exists i_0 : f_{i_0}(x) < f_{i_0}(y) \qquad (2)$$

The first above condition emphasizes that *y* should not be better than *x* at all, while the second above condition indicates that *x* should be better than *y* in at least one aspect (objective function).

#### 3.2.3. Solving an MOP

One of the most considered general approaches for solving an MOP is the Pareto-based approach, in which each objective function of an MOP is treated separately. The Pareto-based approach does not transform a multi-objective problem into single-objective ones for solving an MOP. The output of the Pareto-based approach is a set of non-dominated solutions.

Consider that *X* is a set of generated solutions. It can be said that $x^* \in X$ is a non-dominated solution if there does not exist another solution like *x* such that *x* dominates $x^*$ ( $x \prec x^*$ ). The non-dominated set of solutions or the Pareto optimal set of solutions can be represented as follows:

$$p^* = \{x^* \in X \mid \neg \exists x \in X, x \prec x^*\} \qquad (3)$$

As previously mentioned, the output of the Pareto-based approach is not a unique solution but a set of non-dominated solutions (a Pareto optimal set of solutions). It is the advantage of the Pareto-based approach that provides multiple candidate solutions for a decision maker, and it is his/her responsibility to select the best compromise solution among the set of candidate non-dominated solutions, which are considered to be equally optimal [29]. The selection is essentially a trade-off of one complete solution *x* over another in multi-objective space [7].

## 4. Proposed method

As previously mentioned, this paper aims to detect the overlapping communities of a social network with node attributes in which members share similar attributes, and have dense connections. This goal is achieved by proposing a Multi-Objective BBO-based Overlapping Community Detection algorithm (MOBBO-OCD), which finds overlapping communities of a social network, in which node attributes are available, with considering the two aspects of topological structure and node attributes of the network. Since MOBBO-OCD uses the Pareto-based approach, its final output is a set of non-dominated solutions (partitions) of its input social network. The pseudo code of MOBBO-OCD is shown in Algorithm 1.

**Algorithm 1:** MOBBO-OCD

**Input:** A network with node attributes ($AN=<N,A,E>$)
**Output:** A set of non-dominated partitions of *AN*

1. **Begin**
**%Parameter Initialization%**
2. Initialize *nSIV* (number of nodes of *AN*)
3. Initialize *nHabitat* (number of habitats or size of population)
4. Initialize the immigration rates $\lambda$ (an arithmetic progression from 0 to 1 with common difference of $d=1/(nHabitat-1)$ )
5. Initialize the emmigration rates $\mu$ ( $\mu=1-\lambda$ )
6. Initialize *pMutation* (probability of mutation) according to Eq. (4)
7. Initialize *OVSet* (set of candidate overlapping nodes) according to **Algorithm 2**
**%End of Parameter Initialization%**
**%Initialization%**
8. Generate initial habitats ($H_1,\ldots,H_{nHabitat}$) according to **Algorithm 3**, and store them in *HBT* set
9. Compute HSI values of the habitats of *HBT* based on the objective functions described in Sub-step 2.2
10. Sort *HBT* according to **Algorithm 4**
**%End of Initialization%**
**%Main Loop%**
11. **While** not *T* %*T* is a termination criterion
12. *newHBT* $\leftarrow$ *HBT*
13. **For** each habitat like $H_i$ in *newHBT*
14. **For** each SIV *k* in $H_i$
15. Perform migration according to **Algorithm 5**
16. **If** *rand* $\leq$ *pMutation*
17. Perform the first phase of mutation strategy according to **Algorithm 6**
18. Perform the second phase of mutation strategy according to **Algorithm 7**
19. **End If**
20. **End For**
21. Perform crossover according to **Algorithm 8**
22. Perform the decoding stages described in Sub-step 3.6
23. Compute HSI values of $H_i$ based on the objective functions described in Sub-step 2.2
24. **End For**
25. *HBT* $\leftarrow$ *HBT* $\cup$ *newHBT* %merging *HBT* and *newHBT*
26. Sort *HBT* according to **Algorithm 4**
27. *HBT* $\leftarrow$ Select the first *nHabitat* number of habitats from *HBT*
28. Sort *HBT* according to **Algorithm 4**
29. **End while**
**%End of Main Loop%**
30. **Return** *HBT* as the final output of MOBBO-OCD
31. **End**

According to Algorithm 1, the input of MOBBO-OCD is a social network with node attributes $AN=<N,A,E>$, where *N*, *A* and *E* represent nodes of the network, the corresponding attribute values of the nodes and the edges of the network, respectively. As a matter of fact, *AN* contains the adjacency matrix and nodes' attributes of the social network. With considering Algorithm 1, the strategy of MOBBO-OCD can be outlined in the following 7 steps:

**Step 1: Parameter Initialization.** In the first step of MOBBO-OCD, the parameters of the algorithm are initialized as follows:

**Sub-step 1.1: Initializing *nSIV*.** The first parameter to be initialized is *nSIV*, which represents the number of nodes of *AN* (the input social network).

**Sub-step 1.2: Initializing *nHabitat*.** The second parameter to be initialized is *nHabitat*, which represents the number of habitats or the size of the population.

**Sub-step 1.3: Initializing immigration rates ( $\lambda$ ).** The third parameter to be initialized is immigration rates ( $\lambda$ ). In MOBBO-OCD, the immigration rates are considered to be an arithmetic progression from 0 to 1 with common difference of $d=1/(nHabitat-1)$ .

**Sub-step 1.4: Initializing emigration rates ( $\mu$ ).** The fourth parameter to be initialized is emigration rates ( $\mu$ ). In MOBBO-OCD, the emigration rates are considered to be $\mu=1-\lambda$ . Here, we give an example for better clarifying the performance of immigration rates and emigration rates in MOBBO-OCD. Consider the size of the population (*nHabitat*) to be 5. In this condition, the immigration rates will be 0, 0.25, 0.5, 0.75, 1, respectively, while the emigration rates will be 1, 0.75, 0.5, 0.25, 0, respectively. It can be concluded that, the sum of immigration and emigration rates for each habitat of a population is considered to be 1. Also, with considering the population to be sorted, the first member (the best habitat) of the population has the lowest immigration rate (0) and the highest emigration rate (1) among other members of the population, while the last member (the worst habitat) of the population has the highest immigration rate (1) and the lowest emigration rate (0) among other members of the population.

**Sub-step 1.5: Initializing *pMutation*.** The fifth parameter of MOBBO-OCD to be initialized is *pMutation*, which indicates the probability of mutation in the algorithm. Since different mutation rates will dramatically affect the results of an evolutionary algorithm, in MOBBO-OCD, the value of *pMutation* is calculated based on a method proposed in [29], which gives a good approximation of a mutation probability. In this method, first, a suitable mutation rate for a fixed number of SIVs should be approximated. In our experiments, same as [29], we found that MOBBO-OCD with *pMutation* of 0.1 has a good performance to solve a problem of overlapping community detection in which *nSIV* (number of nodes) is equal to 100. This approximation is conducted with try and error. Then, *pMutation* for other problems with different number of SIVs can be approximated using Eq. (4) [29]:

$$0.1 \times 100 = pMutation \times nSIV \Rightarrow \text{pMutation} = \frac{10}{\text{nSIV}} \quad (4)$$

According to Eq. (4), the value of *pMutation* depends on *nSIV* (number of nodes) of a community detection problem, but the product of *pMutation* and *nSIV* is considered to be a fixed number (10 in our experiments). It can be concluded that with using Eq. (4), *pMutation*s of different community detection problems are made to be dependent on their number of nodes rather than being a fixed number.

**Sub-step 1.6: Initializing *OVSet*.** The last parameter to be initialized in MOBBO-OCD is *OVSet*, which includes the candidate overlapping nodes of *AN*. The members of *OVSet* are recognized by using a method proposed in [46]. This method finds the candidate overlapping nodes based on the following two observations [46]:

- "Observation 1: For each overlapping node of several communities, there usually exists one neighboring node in each community which is densely connected to the overlapping node.
- Observation 2: The links between communities that have at least an overlapping node are spare enough to make these communities unable to form one community."

With considering $n_i$ to be a node in the graph *G* of a social network and based on the two above observations, the following definitions are given in [46]:

- "Definition 1 (Key Neighboring Node): Key Neighboring Node of $n_i$, denoted as $n_i^{KN}$, is the node in the neighborhood of $n_i$ which has the largest number of common neighboring nodes with $n_i$.
- Definition 2 (Key Neighboring Sub-graph): Key Neighboring Sub-graph of $n_i$, denoted as $G_i^{KN}$, is the sub-graph consisting of Key Neighboring Node of $n_i$ ( $n_i^{KN}$ ) and common neighboring nodes of $n_i$ and $n_i^{KN}$ ."

With considering the above definitions and observations, it can be concluded that the node $n_i$ would have a high probability of being an overlapping node if the following two conditions are both satisfied [46]:

- "Condition 1: There should be at least two different Key Neighboring Sub-graphs of $n_i$ in the considered social network.
- Condition 2: The links between any two Key Neighboring Sub-graphs of $n_i$ should be spare."

In order to consider the second condition, Eq. (5) is used to measure the link closeness (*LC*) between two neighboring sub-graphs $G_1^{KN}$ and $G_2^{KN}$ of *G* [46]:

$$LC(G_1^{KN}, G_2^{KN}) = \max\left\{\frac{L(G_1^{KN}, G_2^{KN})}{L(G_1^{KN}, G_1^{KN})}, \frac{L(G_1^{KN}, G_2^{KN})}{L(G_2^{KN}, G_2^{KN})}\right\} \quad (5)$$

Where $L(G_1^{KN}, G_2^{KN})$ denotes the number of links between $G_1^{KN}$ and $G_2^{KN}$, and is defined as $L(G_1^{KN}, G_2^{KN}) = \sum_{i \in G_1^{KN}, j \in G_2^{KN}} A_{ij}$ where *A* is the adjacency matrix of the social network *G*. The links between $G_1^{KN}$ and $G_2^{KN}$ are considered to be spare if their *LC* is equal to or smaller than a given threshold [46]. The same as [46], in MOBBO-OCD, this threshold is considered to be 0.1.

Fig. 1 shows an example for better clarifying the process of finding candidate overlapping nodes of a social network in MOBBO-OCD. Fig. 1(a) shows the graph *G* of a social network with 5 nodes and 6 edges. We want to find out whether node 3 is a candidate overlapping node of *G* or not. For this reason, first, we find the neighboring nodes of node 3, which are node 1, node 2, node 4 and node 5. Then, we find the common neighboring nodes of node 3 and its neighboring nodes. The common neighboring nodes of node 3 and node 1, node 3 and node 2, node 3 and node 4, and node 3 and node 5 are node 2, node 1, node 5 and node 4, respectively. Since the number of common neighboring nodes of node 3 and all of its neighboring nodes are equal to 1, the Key Neighboring Node of node 3 can be any of its neighboring nodes. With choosing node 1 to be the first Key Neighboring Node of node 3, the first Key Neighboring Sub-graph of node 3 ( $G_1^{KN}$ ) is formed, which includes node 1 and node 2. This Key Neighboring Sub-graph is shown in Fig. 1(b). After removing the members of $G_1^{KN}$ (node 1 and node 2) and their related edges from *G*, the reduced sub-graph of *G*, which is shown in Fig. 1(c), is formed. Now, with choosing node 4, from the reduced sub-graph of *G*, as the second Key Neighboring Node of node 3, the second Key Neighboring Sub-graph of node 3 ( $G_2^{KN}$ ) is formed, which includes node 4 and node 5. This Key Neighboring Sub-graph is shown in Fig. 1(d). Since two Key Neighboring Sub-graphs are found for node 3, the first condition is satisfied. If we consider the two Key Neighboring Sub-graphs of node 3, which are shown in Fig. 1(b) and Fig. 1(d), we can see that there is no link between these sub-graphs, which means $LC(G_1^{KN}, G_2^{KN}) = \max\left\{0/1, 0/1\right\} = 0$. If we consider the threshold to be 0.1, the links between the two Key Neighboring Sub-graphs of node 3 will be considered to be spare ( $LC(G_1^{KN}, G_2^{KN}) \le 0.1$ ). Thus, in this condition, node 3 is considered to be a candidate overlapping node of *G*.

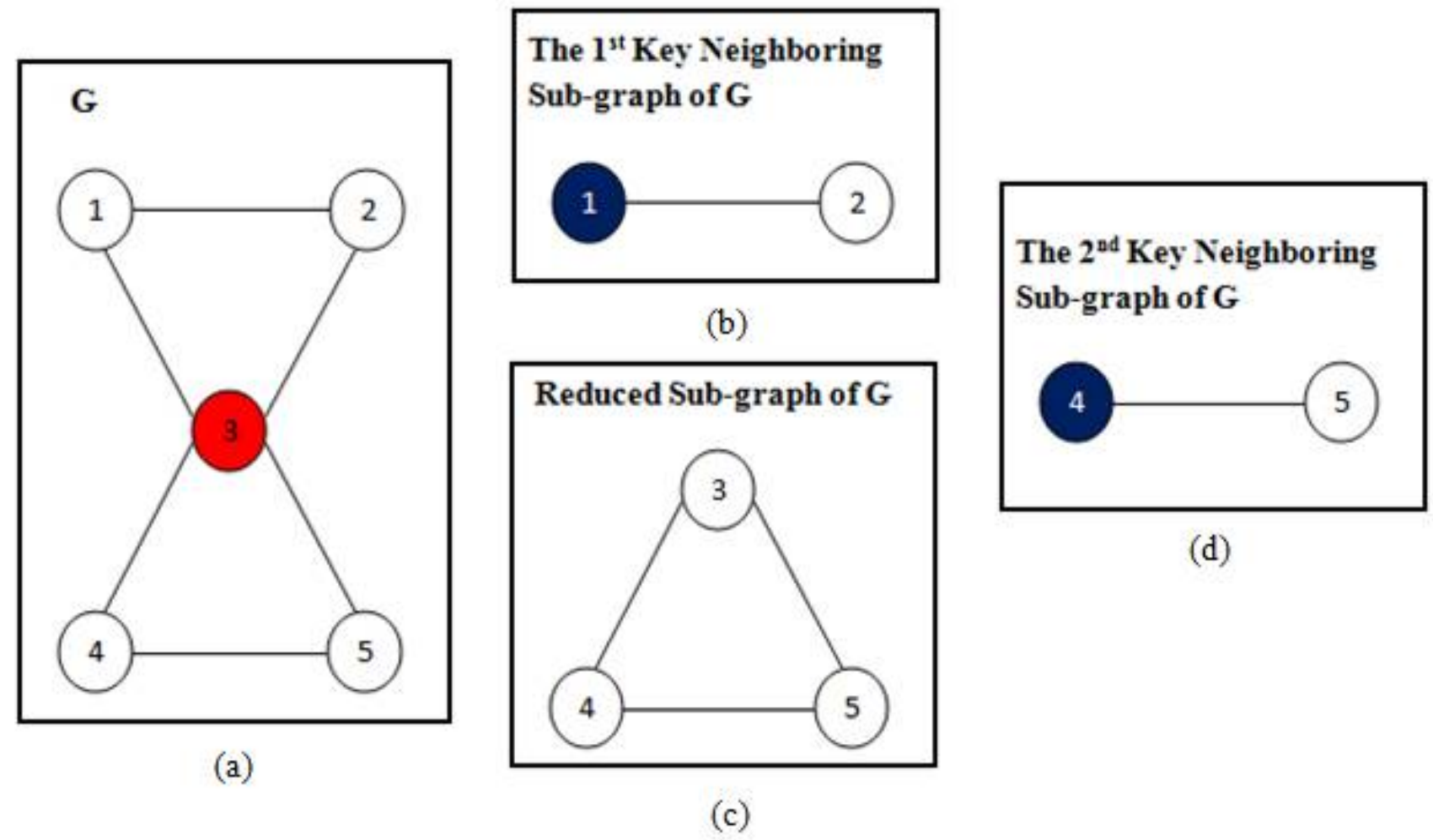

**Fig. 1.** The process of determining whether node 3 is a candidate overlapping node of the graph *G* or not. (a) The graph *G* of a social network with 5 nodes and 6 edges. (b) Node 1 is chosen to be the first Key Neighboring Node of node 3, and the first Key Neighboring Sub-graph of node 3 is formed which includes node 1 and node 2. (c) The obtained reduced sub-graph after removing the first Key Neighboring Sub-graph along with its related edges from *G*. (d) Node 4 is chosen to be the second Key Neighboring Node of node 3, and the second Key Neighboring Sub-graph of node 3 is formed from the reduced sub-graph which includes node 4 and node 5.

If we perform the explained process for other nodes in the network of Fig. 1(a), only one key neighboring sub-graph will be found for each of these nodes, the first condition will not be satisfied for the nodes, and none of them can be a candidate overlapping node of *G*. Thus, it can be concluded that node 3 is the only candidate overlapping node in the network of Fig. 1(a). Algorithm 2 summarizes the process of finding candidate overlapping nodes of a social network in MOBBO-OCD.

**Algorithm 2:** Finding *OVSet* (*AN*, *nSIV*)

1. **Begin**
2. *OVSet*={ }
3. **For** *i*=1 to *nSIV* do
4. *Neighbors* ⟵ Find all neighboring nodes of node *i* [from the adjacency matrix of *AN*]
5. $KNSubgraph_1$={ } %$KNSubgraph_1$ is the first Key Neighboring Sub-graph
6. $KNSubgraph_2$={ } %$KNSubgraph_2$ is the second Key Neighboring Sub-graph
7. **For** *j*=1 to 2 do
8. *CandidateKNNodes* ⟵ Find all neighbors from *Neighbors* which have the most common number of neighboring nodes with node *i* (candidate Key Neighboring Nodes)
9. **If** the number of members in *CandidateKNNodes* is more than 1
10. *KNNode* ⟵ Randomly select one member of *CandidateKNNodes*
11. **Else**
12. *KNNode* ⟵ *CandidateKNNodes*
13. **End If**
14. *CommonNNodes* ⟵ Find all common neighboring nodes of node *i* and *KNNode*
15. $KNSubgraph_j$ ⟵ *KNNode* ∪ *CommonNNodes*
16. *Neighbors* ⟵ *Neighbors*-$KNSubgraph_j$
17. **If** *j*=1 AND *Neighbors* is empty
18. *i* ⟵ *i*+1 And go to line 3
19. **End If**
20. **End For**
21. *LC* ⟵ Calculate the link closeness between *KNSubgraph1* and *KNSubgraph2* according to Eq. (5)
22. **If** *LC* ≤ 0.1
23. *OVSet* ⟵ *OVSet* ∪ *i*
24. **End If**
25. **End For**
26. **Return** *OVSet*
27. **End**

**Step 2: Initialization.** The second step of MOBBO-OCD is related to the process of initializing population. In MOBBO-OCD, the population consists of *nHabitat* number of habitats. Each habitat represents a partition of *AN*. The main parts of the initialization step of MOBBO-OCD are outlined in the following 3 sub-steps:

**Sub-step 2.1: Generating initial habitats.** In this sub-step, the initial habitats are generated, and are stored in *HBT* set. Most of the researches, in which evolutionary algorithms were employed to solve community detection problems, used locus-based adjacency representation for encoding and decoding individuals in populations. The locus-based adjacency representation, which was used in [12], can only encode and decode non-overlapping communities. Since MOBBO-OCD is proposed to detect overlapping communities, we present an extended locus-based adjacency representation, which is called Overlapping Locus-based Adjacency Representation (OLAR), to encode and decode overlapping communities. OLAR performs in 4 stages, which are Encoding, First Decoding, Marking and Final Decoding. The original locus-based adjacency representation only has the first two stages. It means that OLAR and the original locus-based adjacency representation have the same performance in Encoding and First Decoding stages. The difference between the two representations is related to the third stage (Marking) and the fourth stage (Final Decoding), which are introduced in OLAR. It should be noted that, the order of the execution of the second and the third stages of OLAR can be reversed. It means that it is possible to conduct the third stage of OLAR (Marking) before its second stage (First Decoding). The pseudo code of OLAR is shown in Algorithm 3.

**Algorithm 3:** OLAR (*AN*, *nSIV*, *nHabitat*, *OVSet*)

```
1.  Begin
2.    For i=1 to nHabitat do
3.      For j=1 to nSIV do
          %Encoding Stage%
4.        Neighbors ⟵ Find all neighboring nodes of node j [from the adjacency matrix of AN]
5.        Hi.SIV(j) ⟵ Randomly select one neighbor from Neighbors as the value of the j^th SIV of Hi
          %End of Encoding stage%
          %Marking Stage%
6.        If node j exists in OVSet
7.          Hi.Status(j) ⟵ Randomly select one of the numbers (0 or 1) as the status of the j^th SIV of Hi
8.        Else
9.          Hi.Status(j)=0
10.       End If
          %End of Marking Stage%
11.     End For
12.     For j=1 to nSIV do
          %First Decoding Stage%
13.       Hi.Community(j) ⟵ Find the non-overlapping community of j^th SIV (node j) of Hi
          %End of First Decoding Stage%
14.     End For
15.     For j=1 to nSIV do
          %Final Decoding Stage%
16.         Hi.FinalCommunity(j) ⟵ Find the overlapping communities of j^th SIV (node j) of Hi with considering Hi.Status(j) and Hi.Community(j)
          %End of Final Decoding Stage%
17.     End For
18.   End For
19.   Return H1,H2,...,HnHabitat
20. End
```

The difference between OLAR and the original locus-based adjacency representation along with the process of generating a habitat with OLAR is explained with an example, illustrated in Fig. 2. Fig. 2(a) shows the graph *G* of a social network with 5 nodes and 6 edges. We want to generate the habitat *H* with OLAR.

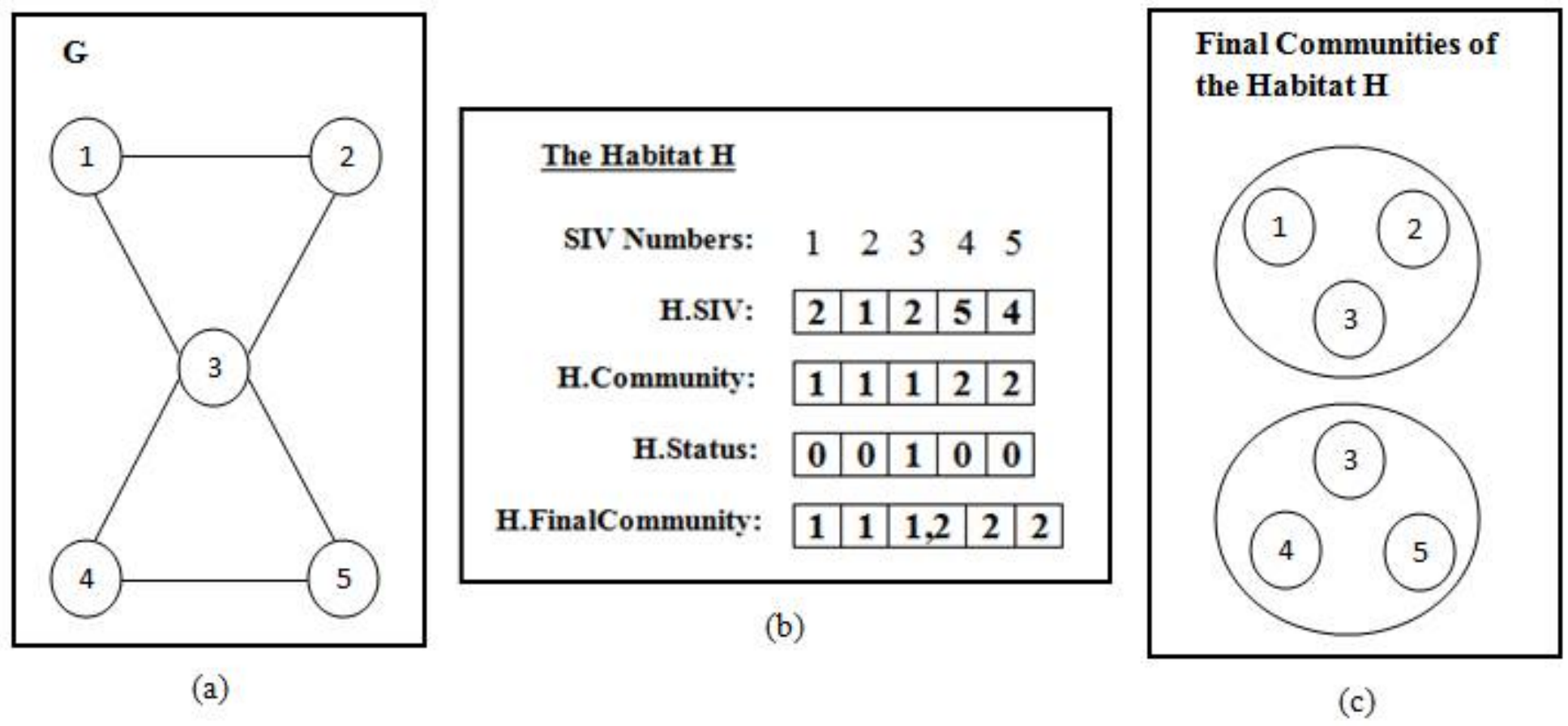

**Fig. 2.** The process of generating a habitat with the proposed overlapping locus-based adjacency representation (OLAR). (a) The graph *G* of a social network with 5 nodes and 6 edges. (b) The habitat *H* with the values of its SIVs (*H.SIV*), the non-overlapping community labels of its SIVs (*H.Community*), the overlapping or non-overlapping statuses of its SIVs (*H.Status*) and the final community labels of its SIVs (*H.FinalCommunity*). (c) The final communities of the habitat *H*.

In OLAR, each habitat (like *H*) deals with *nSIV* variables (SIVs), where *nSIV* represents the number of nodes in *AN*. In the first stage of OLAR (Encoding), for each SIV in *H*, a value is randomly chosen from a set of possible values which includes the neighboring nodes of that SIV. The set of possible values (neighboring nodes) of each SIV can be inferred from the adjacency matrix of *AN*. According to Fig. 2(b), the habitat *H* deals with 5 SIVs because there are 5 nodes in *G* (see Fig. 2(a)). Each of these 5 SIVs is related to a node in the network, and has a set of possible values (neighboring nodes). For example, according to Fig. 2(a), since node 3 has connections with node 1, node 2, node 4 and node 5, the set of possible values of the third SIV of the habitat *H* is {1,2,4,5}. Also, according to Fig. 2(a), the set of possible values of the first SIV of the habitat *H* is {2,3}, the set of possible values of the second SIV of the habitat *H* is {1,3}, the set of possible values of the fourth SIV of the habitat *H* is {3,5}, and the set of possible values of the fifth SIV of the habitat *H* is {3,4}. According to Fig. 2(b), *H.SIV* is the output of the first stage of OLAR (Encoding) with considering the habitat *H*. This output represents a possible encoding of the habitat *H* by OLAR. *H.SIV* contains the corresponding value of each SIV of the habitat *H*, which is randomly chosen from the set of possible values (neighboring nodes) of each SIV. For example, according to *H.SIV* in Fig. 2(b), the corresponding value of the second SIV is 1, which is randomly chosen from the set of possible values of the second SIV ({1,3}). It can be concluded that, in the first stage of OLAR (Encoding), SIVs and their possible values are the nodes of *AN*. Also, if a value *j* is assigned to the $i^{th}$ SIV of a habitat, it is interpreted as an edge between node *i* and node *j* in *AN*. It means that, in the partition of *AN*, which is generated by that habitat, node *i* and node *j* are in the same community [2].

After encoding the habitat *H*, First Decoding stage (the second stage) of OLAR is performed to identify all the non-overlapping communities which are generated by the habitat *H*. In First Decoding stage, SIVs and their related values in *H.SIV* are considered in order to assign the SIVs (nodes), which are part of the same component, to one community. According to Fig. 2(b), *H.Community* is the output of First Decoding stage of OLAR with considering the habitat *H*. *H.Community* contains the non-overlapping community label of each SIV of the habitat *H*. According to *H.SIV* of Fig. 2(b) and the graph *G* of Fig. 2(a), since node 1, node 2 and node 3 are parts of the same component, their community labels are the same in *H.Community*. Also, node 4 and node 5 are parts of the same component, and their community labels are the same in *H.Community*.

So far, the performances of OLAR and the original locus-based adjacency representation are the same. Thus, it can be concluded that both representations are able to automatically determine the number of communities by recognizing the components of each habitat. But at the end of First Decoding stage, the process of the original locus-based adjacency representation is finished. Thus, the original representation is unable to find overlapping communities.

In the third stage of OLAR (Marking), we mark the statuses of the non-overlapping SIVs (nodes) of a habitat with 0 and the statuses of the overlapping SIVs of a habitat with 1. It was discussed in Sub-step 1.6 that *OVSet* contains the candidate overlapping nodes. Thus, the non-overlapping nodes are the nodes from *AN*, which do not exist in *OVSet*. Now, we want to perform Marking stage of OLAR for the habitat *H* of Fig. 2(b). From Sub-step 1.6 and Fig. 1, we know that the related *OVSet* of the graph *G* represented in Fig. 2(a) (which is the same as the graph *G* represented in Fig. 1(a)) contains just one member, which is node 3. Thus, the other nodes of the network (node 1, node 2, node 4 and node 5) are non-overlapping nodes, and the statuses of their related SIVs are marked with 0 in the habitat *H*. On the other hand, in Marking stage, it is randomly determined that a candidate overlapping node is considered to be overlapping (1) or non-overlapping (0). Thus, since node 3 is a candidate overlapping node (see Fig. 1), its status (overlapping or non-overlapping) in *H* is randomly determined. According to Fig. 2(b), *H.Status* is the output of Marking stage of OLAR with considering the habitat *H*. In *H.Status*, the first, the second, the fourth and the fifth values are 0, because they are related to non-overlapping nodes (nodes 1, 2, 4 and 5). But, the third value of *H.Status* is randomly considered to be 1 because its related node (node 3) is a candidate overlapping node (see Fig. 1).

In the last stage of OLAR (Final Decoding), the final communities of a habitat are extracted. As discussed earlier, in the second stage of OLAR (First Decoding), the non-overlapping communities of the habitat *H* are determined. Also, in the third stage of OLAR (Marking), the overlapping and non-overlapping SIVs (nodes) of the habitat *H* are marked. With considering the results of the second and the third stages, in the last stage of OLAR, the community labels of the non-overlapping SIVs of the habitat *H* (the SIVs with status of 0 in *H.Status*) are considered to be the same as their related community labels in *H.Community*. But for the overlapping SIVs of the habitat *H* (the SIVs with status of 1 in *H.Status*), all of the related community labels of their neighboring nodes are extracted from *H.Community*, and are considered as their new community labels. According to Fig. 2(b), *H.FinalCommunity* is the output of the fourth stage of OLAR with considering the habitat *H*. The first, the second, the fourth and the fifth values of *H.FinalCommunity* are the same as their related values in *H.Community*, because their related values in *H.Status* are 0. Thus, according to *H.FinalCommunity*, the final community labels of the first and the second SIVs (node 1 and node 2) are 1, and the final community labels of the fourth and the fifth SIVs (node 4 and node 5) are 2. On the other hand, the third value of *H.FinalCommunity* is 1,2 (1 and 2). It means that the third node of the Graph *G* of Fig. 2(a), which is an overlapping node (according to the third value of *H.Status*), is a member of community 1 and community 2. The reason is that, according to *H.Community*, some of the neighboring nodes of node 3 (node 1 and node 2) are the members of community 1, and its other neighboring nodes (node 4 and node 5) are the members of community 2. Thus, according to the last stage of OLAR, the third value of *H.FinalCommunity*, which is related to the node 3, should contain the community labels of the neighboring nodes of node 3, which are 1 and 2 in *H.Community*. Fig. 2(c) graphically represents the final communities of the habitat *H*, which are detected in the last stage of OLAR (Final Decoding). According to this figure, node 3 is overlapping, and is a member of the both generated communities. But the other nodes are non-overlapping, and each of them is only the member of one community. It should be considered that since the related values of *H.Status* for candidate overlapping nodes are randomly determined in Marking stage, the third value of *H.Status* could be 0. In this condition, *H.FinalCommunity* would be the same as *H.Community*, and the final communities would be non-overlapping. Thus, OLAR is able to automatically detect non-overlapping communities, too.

It can be concluded that, the main advantage of OLAR over the original locus-based adjacency representation is its ability to detect overlapping communities, automatically. In other words, OLAR is able to automatically find both overlapping and non-overlapping communities, while the original locus-based adjacency representation can only find non-overlapping communities, automatically.

**Sub-step 2.2: Computing HSI values.** After generating initial habitats in Sub-step 2.1, the HSI values of the generated habitats are computed in this sub-step. As previously mentioned, the aim of MOBBO-OCD is to find overlapping communities in which the members have dense connections and share similar attributes. With considering this goal, we choose two objective functions to be maximized in MOBBO-OCD, which are Extended Modularity (*EQ*) [34] and *SimAtt* [29].

Newman [24] introduced a concept called Modularity which measures the communities from the perspective of the topological structure of a network, since it is often used to evaluate whether the division is good in the sense that there are many edges within communities and only a few between them [29, 47]. But the original definition of Modularity is not able to handle overlapping communities. Thus, Shen et al. [34] proposed Extended Modularity (*EQ*) to evaluate the goodness of overlapped community decomposition, which is defined as follows:

$$EQ = \frac{1}{2m} \sum_{i} \sum_{v \in C_i, w \in C_i} \frac{1}{O_v O_w} \left[ A_{vw} - \frac{k_v k_w}{2m} \right] \qquad (6)$$

Where $A$ is the adjacency matrix of the corresponding network. Thus, $A_{vw}$ is 1 if an edge exists between node $v$ and node $w$, and is 0, otherwise. $C_i$ denotes a community ( $1 \le i \le l$ and $l$ is the number of communities), $O_v$ is the number of communities to which node $v$ belongs, $k_v$ is the degree of node $v$, and $m$ is the total number of edges in the corresponding network.

Extended Modularity is considered as the first objective function of MOBBO-OCD, because it considers the first aspect of MOBBO-OCD, which is to find overlapping communities in which the members have dense connections. As a matter of fact, Extended Modularity considers the topological structure of a network in the process of overlapping community detection (the larger the value of Extended Modularity, the better the result of overlapping community detection from the perspective of topological structure). It should be noted that the value of Extended Modularity will be the same as that of the original Modularity [24], when each node of the corresponding network belongs to only one community [34].

Reihanian et al. [29] introduced a metric named *SimAtt* that measures the similarity of attributes of the nodes of a community, and is defined as follows:

$$SimAtt = 1/N_{cm} \cdot \sum_{i=1}^{N_{cm}} \max_{1 \le j_1 \le k_1, 1 \le j_2 \le k_2, \ldots, 1 \le j_m \le k_m} \{ (\sum_{h=1}^{m} n_{ij_h}) / (m \times n_i) \} \qquad (7)$$

Where $N_{cm}$ represents the number of detected communities of the corresponding network, $m$ represents the number of attributes of each node of the network, $n_{ij_h}$ refers to the number of nodes which their $h^{th}$ attribute has the value $j_h$, and belongs to community $i$, $n_i$ refers to the number of nodes in community $i$, $k_1$ is the number of distinct values for the first attribute of nodes, ..., and $k_m$ is the number of distinct values for the $m^{th}$ attribute of nodes [29].

*SimAtt* is considered as the second objective function of MOBBO-OCD, because it considers the second aspect of MOBBO-OCD which is to find overlapping communities in which the members share similar attributes (the larger the value of *SimAtt*, the better the result of overlapping community detection from the perspective of attribute similarity [29]).

At the end of the current sub-step, HSI values are computed for each habitat of the initial population. Since Extended Modularity and *SimAtt* are the two objective functions to be maximized in MOBBO-OCD, HSI includes the related values of the two mentioned objective functions for each habitat of the population.

**Sub-step 2.3: Sorting.** We know that each habitat represents a solution, and the goodness of it can be approximated by the values of its HSI. Thus, in this sub-step, the generated habitats are sorted according to their HSIs. In single-objective BBO, HSI of each habitat includes only one value. Thus, in this condition, the habitats can easily be sorted according to their HSI values. But we know that, the HSI of each generated habitat by MOBBO-OCD, which is a multi-objective algorithm, includes two values, which are the related values of Extended Modularity and *SimAtt* for the habitat. Thus, the sorting strategy of the single-objective BBO cannot be applied to MOBBO-OCD. For this reason, we employed the famous sorting strategy of NSGA-II [8], which is a multi-objective genetic algorithm, as the sorting strategy of MOBBO-OCD. In this sorting strategy, first, the non-dominated sorting is conducted, and a rank is assigned to each solution. Then, a metric, which is called Crowding-distance, is calculated for each solution based on its rank. After that, the habitats are sorted according to their non-dominated ranking and Crowding-distance. As a matter of fact, the sorting strategy, which is used in MOBBO-OCD, follows two goals. First, it aims to reach to a Pareto-optimal set which contains non-dominated solutions. This goal can be achieved by the non-dominated sorting algorithm. Then, the sorting strategy aims to reach to a good spread of solutions, which is the second goal of the sorting strategy, to preserve diversity in the generated set of solutions. This goal can be achieved by using Crowding-distance [29]. Algorithm 4 [29] shows the pseudo code of the sorting process in MOBBO-OCD. For more details about non-dominated sorting and calculating Crowding-distance, please refer to [8].

**Algorithm 4:** Sorting (*HBT*)

1. **Begin**
2. Determine the rank (non-dominated set) of each habitat like $H_i$ in *HBT* according to [8]
3. Determine the Crowding-distance of each habitat like $H_i$ in HBT according to [8]
4. Sort *HBT* according to the Crowding-distance values of its habitats
5. Sort *HBT* according to the non-dominated rankings of its habitats
6. **Return** *HBT*
7. **End**

At the end of Initialization step (step 2) of MOBBO-OCD, the initial habitats are generated, their HSI values are computed, and they are sorted according to their non-dominated ranking and Crowding-distance in a descending order of their performance. Thus, so far the algorithm has the initial sorted population (*HBT* set), which is ready to go through the evolution process in the next step.

**Step 3: Evolution process.** In this step, the evolution process of MOBBO-OCD is conducted for each habitat of the input population (*HBT* set). The main parts of this step are outlined in the following 7 sub-steps:

**Sub-step 3.1: Copying.** In this sub-step, the contents of the input population (*HBT* set) are copied to a new set (*newHBT* set). This new set will contain the final output of step 3.

**Sub-step 3.2: Migration.** In this sub-step, the information sharing process is conducted in MOBBO-OCD by the means of migration operator. As previously mentioned, the migration operator, which was introduced in BBO, is used to change and modify the SIV values of the existing habitats. The migration operator which is used in MOBBO-OCD is rank-based, with this assumption that its input population is sorted. In the sorted population, the solution which is located in the first place has the lowest $\lambda$ value (immigration rate) and the highest $\mu$ value (emigration rate), …., and the solution which is located in the last place has the highest $\lambda$ value and the lowest $\mu$ value [29]. In the rank-based migration operator of MOBBO-OCD, for the habitat $H_i$, first, the algorithm decides whether to accept an SIV value from other habitats of the population or not, according to the immigration rate of $H_i$ ($\lambda_i$). If the immigration is selected, then an emigrating habitat from the population should be chosen. The emigrating habitat $H_j$ is probabilistically selected based on its emigration rate ($\mu_j$). Now, with the selection of $H_j$ as the emigrating habitat, the related SIV value ($k^{th}$ SIV value) of $H_i$ is replaced with that of $H_j$, as follows:

$$H_i.SIV(k) \leftarrow H_j.SIV(k) \qquad (8)$$

The pseudo code of the migration mechanism of MOBBO-OCD is shown in Algorithm 5 [29].

**Algorithm 5:** Migration ($H_i$, $k$, $\mu$ , $\lambda$ , *HBT*)

1. **Begin**

2. Select $H_i$ with probability based on $\lambda_i$
3. **If** $H_i$ is selected
4. Select $H_j$ from *HBT* with probability based on $\mu_i$ (with roulette wheel mechanism)
5. **If** $H_j$ is selected
6. $H_i.SIV(k) \longleftarrow H_j.SIV(k)$
7. **End If**
8. **End If**
9. **Return** $H_i$
10. **End**

**Sub-step 3.3: Mutation – Phase 1.** In this sub-step, the first phase of the mutation strategy of MOBBO-OCD is explained. As the mutation strategy in BBO suggests [35], in this phase, some probabilistically selected SIV values of each habitat of the population are changed. MOBBO-OCD employs the two method mutation strategy (for unweighted networks) of [29] as its first phase of mutation strategy. This two method mutation strategy helps the algorithm to escape from local optima. The *pMutation*, which was discussed in Sub-step 1.5, controls the chance of mutation for each habitat of the population.

In the first phase of the mutation strategy of MOBBO-OCD, if the value of the $k^{th}$ SIV of the habitat *H* (*H.SIV(k)*) is chosen to be mutated, one of the following methods of mutation is selected randomly [29]:

- First method [29, 49]: The aim of this method is to place node *k* into a community where most of its neighboring nodes are. Thus, first the neighboring nodes of this node are identified in the habitat *H*, and their community labels are retrieved. Then, the community label, which most of them are belong to, is found. Finally, one of the neighboring nodes of node *k*, which have this community label, is randomly chosen to be placed as a new $k^{th}$ SIV value of the habitat *H*. This method considers the structure of a habitat (partition) to mutate SIV values [29].
- Second method [29]: This method considers the contents of a population for mutating SIV values. In this method, first, the most frequent value of the $k^{th}$ SIV in all of the members of the population along with the $k^{th}$ SIV value of the first member of the population (the best habitat) are found. After that, the value of the $k^{th}$ SIV of the habitat *H* is compared with the mentioned retrieved values. If this value is not equal to the most frequent value, then it is replaced with it. But, if this value is equal to the most frequent value but is not equal to the $k^{th}$ value of the first habitat of the population, then it is replaced with the latter. But if the value is equal to both of the mentioned values (this will happen when the $k^{th}$ SIV value of the best member of the population is the most frequent $k^{th}$ SIV value of the population), we randomly select a neighboring node of node *k*, which is not equal to the $k^{th}$ SIV value of the first habitat in the population, and we place it as a new $k^{th}$ SIV value of the habitat *H* [29].

The pseudo code of the first phase of the mutation strategy of MOBBO-OCD is shown in Algorithm 6 [29].

**Algorithm 6:** Mutation – Phase 1 (*AN*, $H_i$, *k*, *HBT*)

1. **Begin**
2. *rand* $\longleftarrow$ Randomly select a number between 0 and 1
3. **If** *rand* $\leq$ 0.5
4. **%The First Method%**
5. *Neighbors* $\longleftarrow$ Find all neighboring nodes of node *k* of *AN*
6. *l* $\longleftarrow$ Find a community label from $H_i.Community$ that most neighbors from *Neighbors* belong to
7. *r* $\longleftarrow$ Randomly choose one neighbor from *Neighbors* with community label *l*
8. $H_i(j)=r$
9. **Return** $H_i$
10. **%End of the First Method%**
11. **Else**
12. **%The Second Method%**
13. *Pos1* $\longleftarrow$ Find the most frequent $k^{th}$ SIV value in *HBT*
14. *Pos2* $\longleftarrow$ Find $H_1(j)$ %$H_1$ is the best solution of *HBT*
15. **If** $H_i(j)!=Pos1$
16. $H_i(j)=Pos1$ and **Return** $H_i$

17. **End If**
18. **If** $H_i(j)$!=*Pos2*
19. $H_i(j)$=*Pos2* and **Return** $H_i$
20. **End If**
21. **If** ($H_i(j)$=*Pos1* AND $H_i(j)$=*Pos2*)
22. *Neighbors* ⟵ Find all neighboring nodes of node *k* of *AN*
23. *Neighbors2* ⟵ Find neighbors in *Neighbors* that are not equal to *Pos1* (OR *Pos2*)
24. *r* ⟵ Randomly choose one neighbor from *Neighbors2*
25. $H_i(j)$=*r*
26. **End If**
27. **Return** $H_i$
28. **%End of the Second Method%**
29. **End If**
30. **End**

**Sub-step 3.4: Mutation – Phase 2.** In this sub-step, the second phase of the mutation strategy of MOBBO-OCD is explained. According to Sub-step 3.3, the first phase of the mutation strategy tends to change some probabilistically selected SIV values of the habitat *H* in *H.SIV*. But in the second phase, the mutation strategy tends to change some probabilistically selected SIV statuses of the habitat *H* in *H.Status*. In this phase of the mutation strategy of MOBBO-OCD, if the $k^{th}$ status of the habitat *H* (*H.Status(k)*) is chosen to be mutated, first the *OVSet* (please refer to Sub-step 1.6) will be checked to see if node *k* is a candidate overlapping node or not. If node *k* is a candidate overlapping node but its related status in the habitat *H* is 0 (*H.Status(k)*=0), which means this node is considered to be non-overlapping in *H*, its status will change to 1 in order to be considered as an overlapping node. If node *k* is a candidate overlapping node, and its related status in the habitat *H* is 1 (*H.Status(k)*=1), which means this node is considered to be overlapping in *H*, its status will change to 0 in order to be considered as a non-overlapping node. With considering the above explanation, it can be concluded that the second phase of the mutation strategy of MOBBO-OCD considers the (overlapping or non-overlapping) SIV statuses of each habitat of the population for mutation. The pseudo code of this phase is shown in Algorithm 7.

**Algorithm 7:** Mutation – Phase 2 (*k*, *OVSet*, $H_i$)

1. **Begin**
2. **If** node *k* is a member of *OVSet*
3. **If** the status of the $k^{th}$ SIV of $H_i$ ($H_i$.*Status(k)*) is 0 (node *k* is considered to be non-overlapping in $H_i$)
4. Set $H_i$.*Status(k)* to 1 (consider node *k* to be overlapping in $H_i$)
5. **End If**
6. **Else If** the status of the $k^{th}$ SIV of $H_i$ ($H_i$.*Status(k)*) is 1 (node *k* is considered to be overlapping in $H_i$)
7. Set $H_i$.*Status(k)* to 0 (consider node *k* to be non-overlapping in $H_i$)
8. **End If**
9. **Return** $H_i$
10. **End**

**Sub-step 3.5: Crossover.** In this sub-step, MOBBO-OCD employs the double-point crossover operator to share information between the (overlapping or non-overlapping) SIV statuses of the habitats of the population. As a matter of fact, the double-point crossover operator is used in the algorithm to change and modify the SIV statuses of the existing habitats. Like the migration operator, the double-point crossover operator of MOBBO-OCD is rank-based, with this assumption that its input population is sorted. In the rank-based double-point crossover operator of MOBBO-OCD, for the habitat $H_i$, first, the algorithm decides whether to share the SIV statuses of $H_i$ ($H_i$.*Status*) with another habitat of the population or not, according to its immigration rate ( $\lambda_i$ ). If the algorithm decides to share the SIV statuses of $H_i$, then another habitat from the population should be chosen for the sharing process. In this condition, the habitat $H_j$ is probabilistically selected based on its emigration rate ( $\mu_j$ ). Now, with the selection of $H_j$, the status sharing between the two habitats are conducted. For this reason, first, two crossover points ($c_1$ and $c_2$) are randomly selected in the range [1,*nSIV*]. Then, with considering $c_1$ to be smaller than $c_2$, the status vector of $H_i$ is updated in the way that its values in the range ($c_1$+1,$c_2$) are the same as its previous values, and its remained values are the same as the related status values of $H_j$. This process is shown as follows:

$$H_i.Status = [H_j.Status(1, c_1)\ \mathrm{H_i.Status(c_1 + 1, c_2)}\ \mathrm{H_j.Status(c_2 + 1, end)}] \quad (9)$$

The pseudo code of the double-point crossover in MOBBO-OCD is shown in Algorithm 8.

**Algorithm 8:** Double-point Crossover ($H_i$, $\mu$ , $\lambda$ , *nSIV*, *HBT*)

1. **Begin**
2. Select $H_i$ with probability based on $\lambda_i$
3. **If** $H_i$ is selected
4. Select $H_j$ from *HBT* with probability based on $\mu_i$ (with roulette wheel mechanism)
5. **If** $H_j$ is selected
6. Select two unique random numbers between 1 and *nSIV*, as the two crossover points (*c1* and *c2*)
7. **If** $c1<c2$
8. $H_i.Status$=[$H_j.Status(1,c_1)$ $H_i.Status(c_1+1,c_2)$ $H_j.Status(c_2+1,end)$]
9. **Else**
10. $H_i.Status$=[$H_j.Status(1,c_2)$ $H_i.Status(c_2+1,c_1)$ $H_j.Status(c_1+1,end)$]
11. **End If**
12. **End If**
13. **End If**
14. **Return** $H_i$
15. **End**

**Sub-step 3.6: Performing the decoding stages.** After changing and modifying some SIV values (in migration and the first phase of mutation sub-steps) and SIV statuses (in the second phase of mutation and crossover sub-steps) of the habitats of the population, in this sub-step, First Decoding and Final Decoding stages of the introduced OLAR (please refer to Sub-step 2.1) are conducted for each of the habitats to update their identified community labels. In other words, after updating *H.SIV* (in Sub-step 3.2 and Sub-step 3.3) and *H.Status* (in Sub-step 3.4 and Sub-step 3.5) for each habitat (like *H*) of the population, in this sub-step, *H.Community* and *H.FinalCommunity* are updated for these habitats, respectively.

**Sub-step 3.7: Updating HSI values.** This sub-step is the last stage of the evolution process. In this sub-step, HSI values of the evolved habitats of the population are updated, with the same process as the one described in Sub-step 2.2. At the end of this sub-step, *newHBT* contains the population of habitats which is the final output of conducting the evolution process on each habitat of *HBT*.

**Step 4: Merging, sorting and selecting.** In the previous step (evolution process), the habitats of *HBT* are evolved, and are stored in *newHBT*. In this step, first, the habitats of *HBT* and *newHBT* are merged, and the obtained set is stored in *HBT* set. Then, the habitats of *HBT* are sorted, according to their non-dominated ranking and Crowding-distance, in a descending order of their performance. The process of sorting is the same as the one described in Sub-step 2.3. After that, the first *nHabitat* numbers of habitats of *HBT* are selected, and are stored in *HBT* set. The selection mechanism of MOBBO-OCD is the one introduced in [8], which performs as follows:

- If two solutions have different non-domination ranks, the one with the lower rank is selected.
- If two solutions have the same rank, the one with higher Crowding-distance is selected.

For more details about the selection process, please refer to [8].

**Step 5: Final sorting.** In this step, the habits of *HBT* are sorted according to their non-dominated ranking and Crowding-distance in a descending order of their performance. The process of sorting is the same as the one described in Sub-step 2.3.

**Step 6: Termination checking.** In this step, the termination criterion is checked. If it is satisfied, the algorithm goes to next step. Otherwise, the algorithm returns to step 3.

**Step 7: Returning final output.** In this step, MOBBO-OCD is stopped, and *HBT*, which is a set of habitats (partitions of *AN*), is returned as the final output of the algorithm.

## 5. Experiments

In this section, a series of experiments are conducted on 14 real-life data sets with different characteristics to evaluate the performance of MOBBO-OCD by comparing its results with those of 15 relevant community detection algorithms. In the following sub-sections, first, 14 real-life data sets, which are used in the experiments, are described. Then, a performance metric, which is called *alpha_SAEM*, is introduced. After that, the experimental settings are explained. Finally, the three experiments of this research along with their results are presented.

### 5.1. Real-life data sets

As previously mentioned, 14 real-life data sets are used in the experiments. The networks of these data sets can be classified into 5 categories of co-appearance networks, co-purchasing networks, co-rating networks, human social interaction networks and lexical networks. These data sets are described as follows (for more information about the references of the data sets, please refer to [29]):

**1. Anna Karenina data set:** This data set contains the network of characters in the famous novel “Anna Karenina” by Leo Tolstoy. Each node of this network is related to a character in the book. Two characters are connected if they appear in the same scene. The original data set contains the description of the characters. These descriptions are used to find the gender of each character (node) to consider it as the node attribute in the network. The node attribute has two values of “Male” or “Female”. The network of Anna Karenina data set, which contains 138 nodes and 493 edges, can be classified into the category of co-appearance networks [29].

**2. David Copperfield data set:** This data set contains the network of characters in the famous novel “David Copperfield” by Charles Dickens. Each node of this network is related to a character in the book. Two characters are connected if they appear in the same scene. The original data set contains the description of the characters. These descriptions are used to find the gender of each character (node) to consider it as the node attribute in the network. The node attribute has two values of “Male” or “Female”. The network of David Copperfield data set, which contains 87 nodes and 406 edges, can be classified into the category of co-appearance networks [29].

**3. Political Books data set:** This data set has been compiled by Valdis Krebs. This data set contains a network in which nodes represent books about US politics sold by the online bookseller Amazon (http://www.amazon.com), and Edges represent frequent co-purchasing of books by the same buyers. Nodes have been given attribute values of “l”, “n”, or “c” to indicate whether they are “liberal”, “neutral”, or “conservative”, respectively. These alignments were assigned separately by Mark Newman based on a reading of the descriptions and reviews of the books posted on Amazon. The network of Political Books data set, which contains 105 nodes and 441 edges, can be classified into the category of co-purchasing networks [29].

**4. Book-Crossing data set:** This data set was collected by Cai-Nicolas Ziegler from Book-Crossing community (http://www.bookcrossing.com). There are 1149780 ratings in Book-Crossing data set which were attached to about 271379 books by 278858 users. For this data set, the categories of 92 books were extracted from Amazon (http://www.amazon.com). These books are from two general categories of “Fiction” and “Non-Fiction”. The “Fiction” category contains 80 books, while the “Non-Fiction” category contains 12. A network of these books was constructed by drawing an edge between each two books, which were rated by at least one same user. This network, which contains 92 nodes (books) and 2493 edges, can be classified into the category of co-rating networks. Also, in this network, the general category of each book is considered as the node attribute [29].

**5. CIAO data set:** Ciao.co.uk was a product review site in which users shared their opinions about a product by means of rating or commenting. These products were divided into different categories. There are 284086 ratings in CIAO data set which were attached to 105114 products (from 28 categories) by 7375 users. For this data set, a network of products, which are in two categories of “Fashion” and “Office Equipment”, was constructed by drawing an edge between each two products, which were rated by at least one same user. This network, which contains 280 nodes (products) and 2131 edges, can be classified into the category of co-rating networks. Also, in this network, the category of each product is considered as the node attribute [29].

**6. Epinions data set:** Epinions.com was a general consumer review site established in 1999, in which users shared their opinions about a product by means of rating or commenting. These products were divided into different categories. There are 922267 ratings in Epinions data set which were attached to 296277 products (from 27 categories) by 22164 users. For this data set, a network of products, which are in three categories of “Computers & Internet”, “Web Sites & Internet Services” and “Gifts”, was constructed by drawing an edge between each two products, which were rated by at least one same user. This network, which contains 456 nodes (products) and 957 edges, can be classified into the category of co-rating networks. Also, in this network, the category of each product is considered as the node attribute [29].

**7. MovieLens Latest data set:** Grouplens Research Project (http://grouplens.org) at the University of Minnesota has collected and made available the MovieLens data sets from the MovieLens web site (http://movielens.org). The MovieLens Latest data set, which is used in the experiments, is the version of MovieLens data set, which was collected in 2016. There are 105339 ratings and 6117 tags in this data set which were attached to 10328 movies by 668 users [29]. In this data set, the genres of the movies are assigned to them. For this data set, first, all the movies in the genres of “Children” or “War”, which were tagged and rated by at least one user, were retrieved. Then, a network of these movies was constructed by drawing an edge between each two movies, which were tagged and rated by at least one same user. This network, which contains 65 nodes (movies) and 809 edges, can be classified into the category of co-rating networks. 31 out of the 65 movies of the network are in the genre of “Children”, while the 34 other movies of the network are in the genre of “War”. In this network, the genre of each movie is considered as the node attribute.

**8. Movie-Tweetings data set:** Movie-Tweetings is a data set consisting of ratings on movies that were contained in well-structured tweets on Twitter. In our experiments, we used the 10k snapshot of this data set which contains 10000 ratings from 3794 users that were given to 3096 movies. In this data set, the genres of the movies are assigned to them. For this data set, first, all the movies in the genres of “Family” or “Documentary”, which were rated by at least one user, were retrieved. Then, a network of these movies was constructed by drawing an edge between each two movies, which were rated by at least one same user. This network, which contains 182 nodes (movies) and 509 edges, can be classified into the category of co-rating networks. 129 out of the 182 movies of the network are in the genre of “Family”, while the 53 other movies of the network are in the genre of “Documentary”. In this network, the genre of each movie is considered as the node attribute [29].

**9. Football data set:** This data set contains the network of American football games between Division IA colleges during regular season Fall 2000. Nodes of the network represent teams, and edges represent the regular season games between the two teams they connect. Each node has the attribute that indicates to which conference it belongs. The values of this attribute in the network are as follows: 0=“Atlantic Coast”, 1=“Big East”, 2=“Big Ten”, 3=“Big Twelve”, 4=“Conference USA”, 5=“Independents”, 6=“Mid-American”, 7=“Mountain West”, 8=“Pacific Ten”, 9=“Southeastern”, 10=“Sun Belt”, 11=“Western Athletic”. The network of Football data set, which contains 115 nodes and 613 edges, can be classified into the category of human social interaction networks [29].

**10. Primary School data set – Day1:** This data set includes the network related to the first day of study from the Primary School data set. The Primary School data set is part of the study of contact networks in a primary school. The data set comprises two networks of face-to-face proximity between students and teachers. For each day of the study, a daily contact network is provided: nodes are individuals, and edges represent face-to-face interactions. Nodes have two attributes: classname (with 11 different values), which indicates the school class and grade of the corresponding individual, and gender (with 3 different values). Teachers are all assigned to the "Teachers" class. Edges between each two individuals represent a daily contact. The edges have two information: duration, which is the cumulative time spent by A and B in face-to-face proximity, over one day, measured in seconds (multiples of 20 seconds); and count, which is the number of times the A-B contact was established during the school day. In the experiments, the contacts between each two individuals, which were over 100 seconds, were considered for the current data set. Thus, the obtained network of this data set contains 236 nodes and 2197 edges. This network can be classified into the category of human social interaction networks [29].

**11. Primary School data set – Day2:** This data set includes the network related to the second day of study from the Primary School data set. Like the previous data set (Day1), the contacts between each two individuals, which were over 100 seconds, were considered in the experiments for the current data set. Thus, the obtained network of this data set contains 238 nodes and 2419 edges. This network can be classified into the category of human social interaction networks [29].

**12. UK-Faculty data set:** This data set contains the personal friendship network of a faculty of a UK university. This network, which can be classified into the category of human social interaction networks, consists of 81 nodes (individuals) and 577 edges. Each edge of the network represents a friendship connection between each two individuals of the network. The numeric ID of the school affiliation of each individual, which has 4 different values in the network, is stored as the node attribute in the network [29].

**13. Contacts in a Workplace data set (Workplace data set):** This data set contains the temporal network of contacts between individuals measured in an office building in France, from June 24 to July 3, 2013. This network, which can be classified into the category of human social interaction networks, contains 92 nodes (individuals) and 755 edges. Edges of the network represent contacts between individuals. The name of each individual's department in the workplace, which has 5 different values in the network, is considered to be the node attribute in the network [29].

**14. AdjNoun data set**: This data set contains the network of common adjective and noun adjacencies for the famous novel "David Copperfield" by Charles Dickens. Nodes of the network represent the most commonly occurring adjectives and nouns in the book. The value of the node attribute in the network is considered to be 0 for "Adjectives" and 1 for "Nouns". Edges connect any pair of words that occur in adjacent position in the text of the book. The network of this data set, which contains 112 nodes and 425 edges, can be classified into the category of lexical networks [29].

### 5.2. Performance metric

As previously mentioned, the output of MOBBO-OCD is a set of non-dominated solutions (partitions of a network into communities). Reihanian et al. introduced a metric called *alpha_SAM* to determine the best compromise solution among the set of non-dominated solutions achieved by a community detection algorithm, with considering the two aspects of node attributes and linkage structure [29]. Although *alpha_SAM* can make a balance between the similarity of nodes' attributes and the strength of connections in evaluating the goodness of a partition (solution), it can only evaluate the goodness of non-overlapping partitions, and is not able to deal with the overlapping ones.

Since MOBBO-OCD is a community detection algorithm which is able to detect overlapping communities, *alpha_SAM* cannot be employed to evaluate its community detection performance. Thus, with considering the aim of this research, we propose a modified version of *alpha_SAM* called *alpha_SAEM* (alpha_SimAttExtendedModularity), which is able to evaluate the goodness of both overlapping and non-overlapping partitions with considering the two aspects of node attributes and linkage structure. Following the F-score criteria in information retrieval, *alpha_SAEM* is defined as follows:

$$alpha_SAEM = \frac{(1+\alpha^2)(SimAtt \times EQ)}{(\alpha^2 \times SimAtt + EQ)} \qquad (10)$$

Where $\alpha$ is a parameter in the range $[0,\infty)$. The role of $\alpha$ is to adjust the weight of *SimAtt* and Extended Modularity (*EQ*) in *alpha_SAEM*. In the case that the similarity of nodes' attributes and the strength of connections are equally important for us, and we want them to have equal effects on the values of *alpha_SAEM*, $\alpha$ should be set to 1. In this case, *alpha_SAEM* is the harmonic mean of *SimAtt* and Extended Modularity, and is calculated as follows:

$$alpha_SAEM_1 = (1+1^2)(SimAtt \times EQ)/(1^2 \times SimAtt + EQ) = \frac{2SimAtt \times EQ}{SimAtt + EQ} \qquad (11)$$

When the similarity of nodes' attributes is more important for us, and we want them to have more effects on the values of *alpha_SAEM*, we set $\alpha$ to the values less than 1 ( $0 \le \alpha < 1$ ). According to Eq. (12), when $\alpha$ approaches zero, *alpha_SAEM* approaches *SimAtt*:

$$\lim_{\alpha \to 0} alpha_SAEM = \lim_{\alpha \to 0} \frac{(1+\alpha^2)(SimAtt \times EQ)}{(\alpha^2 \times SimAtt + EQ)} = \frac{SimAtt \times EQ}{EQ} = SimAtt \qquad (12)$$

On the other hand, when the link is more important for us, and we want Extended Modularity to have more effect on the values of *alpha_SAEM*, we set $\alpha$ to the values more than 1 ( $\alpha > 1$ ). According to Eq. (13), when $\alpha$ approaches $\infty$ , *alpha_SAEM* approaches the pure value of Extended Modularity:

$$\lim_{\alpha \to \infty} alpha_SAEM = \lim_{\alpha \to \infty} \frac{(1+\alpha^2)(SimAtt \times EQ)}{(\alpha^2 \times SimAtt + EQ)} = \frac{SimAtt \times EQ}{SimAtt} = EQ \qquad (13)$$

Thus, $\alpha$ adjusts the emphasis of the two aspects which are the similarity of nodes' attributes and the strength of connections. As a brief, *alpha_SAEM*, like *alpha_SAM*, can make a balance between the similarity of nodes' attributes and the strength of connections.

By using Extended Modularity in *alpha_SAEM* formula, the metric is able to evaluate the goodness of overlapping partitions. On the other hand, *alpha_SAEM* is able to evaluate the goodness of non-overlapping partitions, too. In this case, the performance of *alpha_SAEM* will be the same as that of *alpha_SAM*, because the value of Extended Modularity will be the same as that of the original Modularity (which is used instead of Extended Modularity in *alpha_SAM*) in this condition.

#### 5.3. Experimental settings

All the experiments of this research are conducted on four computers where the first one has Intel Core 2 Duo 2.20GHz CPU and 2 GB RAM, the second one has Pentium Dual-Core 2.60GHz 1.60GHz CPU and 2 GB RAM, the third one has Intel Xeon 2.93GHz CPU and 4 GB RAM, and the fourth one has AMD Ryzen 7PRO 1700X Eight-Core 3.40GHz CPU and 4 GB RAM. MOBBO-OCD is implemented in MATLAB 8.1.0.604 (R2013a), and its population size (*nHabitat*) and the number of its generations are considered to be 100 in all the experiments. The experimental results for all the algorithms of the experiments are reported by averaging the results of their 10 independent runs on each of the 14 data sets of the experiments. In all the experiments, the isolate communities (the communities with one node) achieved by the algorithms are not considered in calculations.

#### 5.4. First experiment

In the first experiment of this research, the performance of MOBBO-OCD is compared with those of the 8 state-of-the-art overlapping community detection algorithms, i.e., CPM (Clique Percolation Method) [25], COPRA (Community Overlap PRopagation Algorithm) [11],

OSLOM (Order Statistics Local Optimization Method) [17], SLPA (Speaker-listener Label Propagation Algorithm) [39], iLCD (intrinsic Longitudinal Community Detection) [4], AGMFIT (Community Detection by Community-Affiliation Graph Model) [41], BIGCLAM (Cluster Affiliation Model for Big Networks) [42] and CoDA (Communities through Directed Affiliations) [44], on the 14 real-life data sets described in Sub-section 5.1. These algorithms conduct their community detection processes with just considering the graph structures of networks.

In this experiment, for each of the 8 state-of-the-art competitor algorithms of MOBBO-OCD, we used the codes, software or information provided by its corresponding researchers. For COPRA, we executed the algorithm 10 times with considering the value of the parameter $v$, which shows the maximum number of communities per nodes, to be varied from 1 to 10. For AGMFIT, the parameter $e$, which shows the edge probability between the nodes of a network that do not share any community, is considered to be $1/(number\text{ of nodes of the network})^2$, which is suggested in the algorithm's manual. For other algorithms, their default settings are considered. Also, it should be noted that some of the algorithms of the first experiment, generate more than one partition, each time they are executed. For these algorithms, their best partition, based on the values of the performance metric for each of their generated partitions, is considered in each of their executions.

After achieving the results of *SimAtt* and Extended Modularity for each of the algorithms of the experiment, they are used to calculate the values of *alpha_SAEM*. Table 1 presents the mean of the best-of-run *alpha_SAEM* values over independent runs of MOBBO-OCD and the other 8 algorithms of the first experiment on the 14 data sets of the experiments. According to Table 1, the values of *alpha_SAEM* is reported for different values of $\alpha$ (0.5, 1 and 1.5, respectively) in order to evaluate the performances of the algorithms when different strengths for nodes' attributes and link structures are considered. According to Table 1, the three cells which are related to the result of applying CPM to Book-Crossing data set and also, the three cells which are related to the result of applying AGMFIT to MovieLens Latest data set are filled with -, because the two algorithms were not able to detect communities of the mentioned data sets.

"+/=/-" is located in the last row of Table 1 to show that MOBBO-OCD shows better performance on +, equal performance on = and worse performance on - data sets, with considering the different values of $\alpha$, in comparison with its competitors. According to the last row of Table 1, MOBBO-OCD notably outperforms all of its competitors in the first experiment. Also, in Table 1, for each data set, the best performance among all the algorithms in terms of mean *alpha_SAEM* value, with considering the different values of $\alpha$, is highlighted in bold-face. According to the table, when $\alpha$ is equal to 0.5, 1 and 1.5, MOBBO-OCD shows the absolute best performance among its 8 competitors on 13 data sets, 13 data sets and 12 data sets, respectively. This indicates that the performance of MOBBO-OCD is quite superior to the performances of the other algorithms in the first experiment.

**Table 1** The mean of the best-of-run *alpha_SAEM* values, for $\alpha$=0.5, $\alpha$=1 and $\alpha$ =1.5, over independent runs of MOBBO-OCD and the other 8 algorithms of the first experiment on the 14 data sets of the experiments

| Data Sets | alpha_SAEM | Algorithms | | | | | | | | |
|---|---|---|---|---|---|---|---|---|---|---|
| | | CPM | COPRA | OSLOM | SLPA | iLCD | AGMFIT | BIGCLAM | CoDA | MOBBO-OCD |
| Anna Karenina | $\alpha$=0.5 | 0.31866 | 0.15079 | 0.51737 | 0.24996 | 0.42033 | 0.39433 | 0.41322 | 0.15683 | **0.69271** |
| | $\alpha$=1 | 0.17709 | 0.11683 | 0.41304 | 0.16411 | 0.25659 | 0.25826 | 0.26771 | 0.07228 | **0.52205** |
| | $\alpha$=1.5 | 0.13783 | 0.10237 | 0.36577 | 0.13630 | 0.20532 | 0.21189 | 0.21841 | 0.05372 | **0.45620** |
| David Copperfield | $\alpha$=0.5 | 0.43111 | 0.40706 | 0.52249 | 0.07315 | 0.49825 | 0.40524 | 0.46431 | 0.15997 | **0.65528** |
| | $\alpha$=1 | 0.29179 | 0.29626 | 0.44129 | 0.04010 | 0.34976 | 0.28784 | 0.32101 | 0.07517 | **0.49349** |
| | $\alpha$=1.5 | 0.24172 | 0.25600 | 0.40137 | 0.03117 | 0.29366 | 0.24311 | 0.26799 | 0.05611 | **0.43238** |
| Political Books | $\alpha$=0.5 | 0.67721 | 0.70089 | 0.67665 | 0.68338 | 0.54729 | 0.65967 | 0.63503 | 0.21357 | **0.78163** |
| | $\alpha$=1 | 0.55761 | 0.59370 | 0.60003 | 0.59723 | 0.36646 | 0.50831 | 0.44607 | 0.09956 | **0.64311** |
| | $\alpha$=1.5 | 0.50090 | 0.55116 | 0.55953 | 0.55427 | 0.30240 | 0.44316 | 0.37461 | 0.07418 | **0.58493** |
| Book-Crossing | $\alpha$=0.5 | - | 0.00000 | **0.35916** | 0.00000 | 0.00802 | 0.09205 | 0.19647 | 0.00873 | 0.30291 |
| | $\alpha$=1 | - | 0.00000 | **0.19048** | 0.00000 | 0.00322 | 0.03957 | 0.09122 | 0.00351 | 0.15276 |
| | $\alpha$=1.5 | - | 0.00000 | **0.14640** | 0.00000 | 0.00233 | 0.02898 | 0.06791 | 0.00254 | 0.11598 |
| CIAO | $\alpha$=0.5 | 0.77184 | 0.77653 | 0.77056 | 0.81925 | 0.76468 | 0.69605 | 0.75294 | 0.69266 | **0.82544** |
| | $\alpha$=1 | 0.63995 | 0.65141 | 0.64613 | 0.68624 | 0.62490 | 0.56858 | 0.62830 | 0.53452 | **0.68893** |
| | $\alpha$=1.5 | 0.57677 | 0.59142 | 0.58553 | 0.62162 | 0.55936 | 0.50884 | 0.56802 | 0.46736 | **0.62381** |
| Epinions | $\alpha$=0.5 | 0.87173 | 0.87717 | 0.87435 | 0.91057 | 0.87261 | 0.55851 | 0.83199 | 0.84008 | **0.93137** |
| | $\alpha$=1 | 0.82731 | 0.84715 | 0.85546 | 0.90883 | 0.82496 | 0.34568 | 0.78738 | 0.81083 | **0.92225** |
| | $\alpha$=1.5 | 0.80115 | 0.83315 | 0.84377 | 0.90784 | 0.79706 | 0.27782 | 0.76122 | 0.79313 | **0.91888** |
| MovieLens Latest | $\alpha$=0.5 | 0.60174 | 0.65121 | 0.61291 | 0.60839 | 0.61095 | - | 0.55007 | 0.60841 | **0.70809** |
| | $\alpha$=1 | 0.49996 | 0.52397 | 0.50276 | 0.48508 | 0.49949 | - | 0.42171 | 0.49064 | **0.54910** |
| | $\alpha$=1.5 | 0.45106 | 0.46608 | 0.45083 | 0.42997 | 0.44719 | - | 0.36684 | 0.43653 | **0.48231** |
| Movie-Tweetings | $\alpha$=0.5 | 0.78183 | 0.84331 | 0.83643 | 0.88330 | 0.78445 | 0.73881 | 0.79888 | 0.67831 | **0.89339** |
| | $\alpha$=1 | 0.62979 | 0.74907 | 0.74545 | 0.80117 | 0.64144 | 0.57612 | 0.66811 | 0.48510 | **0.80843** |
| | $\alpha$=1.5 | 0.55999 | 0.70119 | 0.69687 | 0.75658 | 0.57433 | 0.50486 | 0.60466 | 0.41065 | **0.76307** |
| Football | $\alpha$=0.5 | 0.81062 | 0.81724 | 0.82404 | 0.71384 | 0.68929 | 0.33191 | 0.63060 | 0.22867 | **0.84888** |
| | $\alpha$=1 | 0.69108 | 0.72047 | 0.72345 | 0.64516 | 0.61698 | 0.28286 | 0.47503 | 0.11809 | **0.73124** |
| | $\alpha$=1.5 | 0.63364 | 0.66981 | 0.67094 | 0.60802 | 0.57811 | 0.26024 | 0.41016 | 0.09017 | **0.67374** |
| Primary School – Day1 | $\alpha$=0.5 | 0.67937 | 0.70593 | 0.70763 | 0.66242 | 0.61188 | 0.48380 | 0.63594 | 0.25806 | **0.78216** |
| | $\alpha$=1 | 0.59430 | 0.68997 | 0.68101 | 0.65958 | 0.54198 | 0.47617 | 0.53323 | 0.13461 | **0.71175** |
| | $\alpha$=1.5 | 0.55014 | 0.68102 | 0.66499 | 0.65799 | 0.50500 | 0.47159 | 0.48320 | 0.10304 | **0.68485** |
| Primary School – Day2 | $\alpha$=0.5 | 0.67037 | 0.71223 | 0.68965 | 0.65505 | 0.64709 | 0.45905 | 0.53278 | 0.27520 | **0.77386** |
| | $\alpha$=1 | 0.58216 | 0.69192 | 0.68199 | 0.64737 | 0.58570 | 0.45346 | 0.39781 | 0.14568 | **0.70531** |
| | $\alpha$=1.5 | 0.55380 | **0.68319** | 0.67725 | 0.64281 | 0.55213 | 0.45040 | 0.34223 | 0.11192 | 0.67779 |
| UK-Faculty | $\alpha$=0.5 | 0.56851 | 0.76978 | 0.76309 | 0.75470 | 0.61714 | 0.45853 | 0.64737 | 0.07965 | **0.79042** |
| | $\alpha$=1 | 0.36992 | 0.59427 | 0.58965 | 0.58076 | 0.45118 | 0.31795 | 0.44104 | 0.03375 | **0.60946** |
| | $\alpha$=1.5 | 0.30224 | 0.51850 | 0.51467 | 0.50629 | 0.38484 | 0.26574 | 0.36622 | 0.02465 | **0.53293** |
| Workplace | $\alpha$=0.5 | 0.62354 | 0.53230 | 0.71198 | 0.07181 | 0.44917 | 0.34912 | 0.53344 | 0.05752 | **0.74052** |
| | $\alpha$=1 | 0.40915 | 0.39617 | 0.53224 | 0.05414 | 0.31556 | 0.20762 | 0.33158 | 0.02417 | **0.54481** |
| | $\alpha$=1.5 | 0.34086 | 0.34042 | 0.45811 | 0.04676 | 0.26502 | 0.16488 | 0.26685 | 0.01762 | **0.46842** |
| AdjNoun | $\alpha$=0.5 | 0.14399 | 0.00000 | 0.00000 | 0.00000 | 0.29884 | 0.31373 | 0.35688 | 0.00250 | **0.51988** |
| | $\alpha$=1 | 0.06606 | 0.00000 | 0.00000 | 0.00000 | 0.16061 | 0.18897 | 0.22799 | 0.00100 | **0.35636** |
| | $\alpha$=1.5 | 0.04904 | 0.00000 | 0.00000 | 0.00000 | 0.12388 | 0.15059 | 0.18513 | 0.00072 | **0.29976** |
| +/=/- | $\alpha$=0.5 | 14/0/0 | 14/0/0 | 13/0/1 | 14/0/0 | 14/0/0 | 14/0/0 | 14/0/0 | 14/0/0 | - |
| | $\alpha$=1 | 14/0/0 | 14/0/0 | 13/0/1 | 14/0/0 | 14/0/0 | 14/0/0 | 14/0/0 | 14/0/0 | - |
| | $\alpha$=1.5 | 14/0/0 | 13/0/1 | 13/0/1 | 14/0/0 | 14/0/0 | 14/0/0 | 14/0/0 | 14/0/0 | - |

The results of the first experiment, shown in Table 1, and the above analysis of the results can demonstrate the best performance of MOBBO-OCD among its 8 competitors in the experiment, but to ensure that the performance of MOBBO-OCD is statistically different from those of the other 8 algorithms, we conducted a statistical significance analysis on the results in Table 1, with considering the different values of $\alpha$, using the Friedman test implemented in IBM SPSS Statistics 22. The analysis shows that there is a statistically significant difference between the performances of the algorithms with $\chi^2(8) = 60.133333$ and $p < 0.01$ for $\alpha = 0.5$, $\chi^2(8) = 63.600000$ and $p < 0.01$ for $\alpha = 1$, $\chi^2(8) = 61.885714$ and $p < 0.01$ for $\alpha = 1.5$. Also, the result of Friedman ranking, which is shown in Table 2, indicates that MOBBO-OCD ranks first among all the algorithms in the experiment, with the mean rank value of 8.93 for $\alpha = 0.5$ and $\alpha = 1$, and 8.86 for $\alpha = 1.5$.

**Table 2** The mean and the final ranks of the 9 algorithms of the first experiment according to the Friedman test, with considering the different values of $\alpha$

| Algorithms | $\alpha$=0.5 | | $\alpha$=1 | | $\alpha$=1.5 | |
|---|---|---|---|---|---|---|
| | Mean Rank | Final Rank | Mean Rank | Final Rank | Mean Rank | Final Rank |
| CPM | 4.64 | 5 | 4.50 | 5 | 4.57 | 5 |
| COPRA | 5.93 | 3 | 6.00 | 3 | 6.07 | 3 |
| OSLOM | 6.64 | 2 | 6.93 | 2 | 6.86 | 2 |
| SLPA | 4.71 | 4 | 4.79 | 4 | 4.79 | 4 |
| iLCD | 4.57 | 6 | 4.43 | 6 | 4.36 | 6 |
| AGMFIT | 3.00 | 8 | 3.14 | 8 | 3.21 | 8 |
| BIGCLAM | 4.50 | 7 | 4.36 | 7 | 4.36 | 6 |
| CoDA | 2.07 | 9 | 1.93 | 9 | 1.93 | 9 |
| MOBBO-OCD | 8.93 | 1 | 8.93 | 1 | 8.86 | 1 |

**Table 3** The $p$-values of the Wilcoxon signed-rank test of the 9 algorithms of the first experiment against each other at the significance level of 0.05, with considering the different values of $\alpha$. The $p$-values below the Bonferroni-corrected critical value are highlighted in bold-face.

| | | CPM | COPRA | OSLOM | SLPA | iLCD | AGMFIT | BIGCLAM | CoDA |
|---|---|---|---|---|---|---|---|---|---|
| **$\alpha$=0.5** | COPRA | 5.10E-01 | - | - | - | - | - | - | - |
| | OSLOM | 1.57E-02 | 7.30E-01 | - | - | - | - | - | - |
| | SLPA | 4.70E-01 | 1.40E-01 | 1.09E-01 | - | - | - | - | - |
| | iLCD | 8.75E-01 | 3.31E-01 | 1.32E-02 | 9.25E-01 | - | - | - | - |
| | AGMFIT | 3.03E-02 | 6.40E-02 | 7.63E-03 | 2.72E-01 | 1.57E-02 | - | - | - |
| | BIGCLAM | 8.26E-01 | 6.38E-01 | 1.57E-02 | 6.38E-01 | 7.78E-01 | 1.52E-03 | - | - |
| | CoDA | 1.89E-03 | 3.51E-03 | **1.23E-03** | 1.10E-02 | **1.23E-03** | 6.40E-02 | 1.89E-03 | - |
| | MOBBO-OCD | **9.82E-04** | **9.82E-04** | 2.87E-03 | **9.82E-04** | **9.82E-04** | **9.82E-04** | **9.82E-04** | **9.82E-04** |
| **$\alpha$=1** | COPRA | 6.40E-02 | - | - | - | - | - | - | - |
| | OSLOM | 3.51E-03 | 4.33E-01 | - | - | - | - | - | - |
| | SLPA | 6.83E-01 | 3.97E-01 | 8.43E-02 | - | - | - | - | - |
| | iLCD | 9.25E-01 | 1.40E-01 | 9.18E-03 | 7.78E-01 | - | - | - | - |
| | AGMFIT | 3.03E-02 | 1.86E-02 | 4.29E-03 | 9.62E-02 | 2.58E-02 | - | - | - |
| | BIGCLAM | 5.10E-01 | 1.40E-01 | 1.10E-02 | 6.38E-01 | 7.78E-01 | 1.32E-02 | - | - |
| | CoDA | **1.23E-03** | 1.89E-03 | **1.23E-03** | 9.18E-03 | **1.23E-03** | 1.09E-01 | 1.89E-03 | - |
| | MOBBO-OCD | **9.82E-04** | **9.82E-04** | 3.51E-03 | **9.82E-04** | **9.82E-04** | **9.82E-04** | **9.82E-04** | **9.82E-04** |
| ***α*=1.5** | COPRA | 3.03E-02 | - | - | - | - | - | - | - |
| | OSLOM | 3.51E-03 | 3.63E-01 | - | - | - | - | - | - |
| | SLPA | 5.10E-01 | 3.63E-01 | 9.62E-02 | - | - | - | - | - |
| | iLCD | 8.26E-01 | 6.40E-02 | 4.29E-03 | 5.51E-01 | - | - | - | - |
| | AGMFIT | 3.55E-02 | 1.32E-02 | 2.33E-03 | 6.40E-02 | 2.58E-02 | - | - | - |
| | BIGCLAM | 3.97E-01 | 6.40E-02 | 9.18E-03 | 5.10E-01 | 5.51E-01 | 3.03E-02 | - | - |
| | CoDA | **1.23E-03** | 1.89E-03 | **1.23E-03** | 7.63E-03 | **1.23E-03** | 1.09E-01 | 2.33E-03 | - |
| | MOBBO-OCD | **9.82E-04** | 1.89E-03 | 4.29E-03 | **9.82E-04** | **9.82E-04** | **9.82E-04** | **9.82E-04** | **9.82E-04** |

We subsequently performed a post-hoc test using the Wilcoxon signed-rank test implemented in IBM SPSS Statistics 22 at the significance level of 0.05, with considering the different values of $\alpha$, to examine where the differences between the performances of the algorithms, that the Friedman test indicates, actually occur. With considering the different values of $\alpha$, Table 3 shows the $p$-values of the 9 algorithms of the first experiment against each other. In Table 3, the $p$-values below the Bonferroni-corrected critical value, which indicate the significant differences between the performances of the corresponding algorithms, are highlighted in bold-face. We used the Bonferroni correction on the results of Wilcoxon signed-rank test because multiple comparisons are conducted in this experiment, which makes it more likely that a result is declared significant when it is not [1]. According to Table 3, compared with the other 8 algorithms, MOBBO-OCD obtains the largest number of $p$-values lower than the significance level with considering the Bonferroni correction, for all different values of $\alpha$. This indicates that the performance of MOBBO-OCD is significantly different from those of the most of the other algorithms of the experiment.

### 5.5. Second experiment

In the second experiment of this research, the performance of MOBBO-OCD is compared with those of the 5 community detection algorithms, i.e., EM-BBO (Extended Modularity maximization BBO algorithm), SimAtt-BBO (*SimAtt* maximization BBO algorithm) [29], CESNA (Communities from Edge Structure and Node Attributes) [43], topic-oriented community detection algorithm [31, 47] and OV-SimAtt-BBO (OVerlapping *SimAtt* maximization BBO algorithm), on the 14 real-life data sets described in Sub-section 5.1. EM-BBO (which is based on BBO) is an overlapping community detection algorithm proposed in this paper, which uses OLAR and the parameter settings of the Modularity maximization BBO algorithm proposed in [49], and tries to maximize Extended Modularity. SimAtt-BBO (which is based on BBO, and tries to maximize *SimAtt*), CESNA, topic-oriented community detection algorithm and OV-SimAtt-BBO (the overlapping version of SimAtt-BBO, which uses OLAR, and is proposed in this paper) consider both the graph structure and the contents of a network in their processes of community detection.

Like the first experiment, for each of the 5 competitor algorithms of MOBBO-OCD in this experiment, we used the codes, software or information provided by its corresponding researchers. In order to have fair comparison, the population size (*nHabitat*) and the number of generations for EM-BBO and SimAtt-BBO have been considered the same as MOBBO-OCD since the mentioned algorithms are evolutionary, too.

Table 4 presents the mean of the best-of-run *alpha_SAEM* values, for $\alpha$=0.5, $\alpha$=1 and $\alpha$=1.5, over independent runs of MOBBO-OCD and the other 5 algorithms of the second experiment on the 14 data sets of the experiments. According to the last row of Table 4, MOBBO-OCD notably outperforms all of its competitors in the second experiment. According to Table 4, when $\alpha$ is equal to 0.5, 1 and 1.5, MOBBO-OCD shows the absolute best performance among its 5 competitors on 11 data sets, 13 data sets and 13 data sets, respectively. This indicates that the performance of MOBBO-OCD is quite superior to the performances of the other algorithms in the second experiment.

The results of the second experiment, shown in Table 4, and the above analysis of the results can demonstrate the best performance of MOBBO-OCD among its 5 competitors in the experiment, but to ensure that the performance of MOBBO-OCD is statistically different from those of the other 5 algorithms, we conducted a statistical significance analysis on the results in Table 4, with considering the different values of *α*, using the Friedman test implemented in IBM SPSS Statistics 22. The analysis shows that there is a statistically significant difference between the performances of the algorithms with $\chi^2(5) = 55.755102$ and $p < 0.01$ for $\alpha = 0.5$, $\chi^2(5) = 59.836735$ and $p < 0.01$ for $\alpha = 1$, $\chi^2(5) = 63.183673$ and $p < 0.01$ for $\alpha = 1.5$. Also, the result of Friedman ranking, which is shown in Table 5, indicates that MOBBO-OCD ranks first among all the algorithms in the experiment, with the mean rank value of 5.79 for $\alpha = 0.5$, and 5.93 for $\alpha = 1$ and $\alpha = 1.5$.

**Table 4** The mean of the best-of-run *alpha_SAEM* values, for *α*=0.5, *α*=1 and *α* =1.5, over independent runs of MOBBO-OCD and the other 5 algorithms of the second experiment on the 14 data sets of the experiments

| Data Sets | alpha_SAEM | Algorithms | | | | | |
|---|---|---|---|---|---|---|---|
| | | EM-BBO | SimAtt-BBO | CESNA | Topic-oriented | OV-SimAtt-BBO | MOBBO-OCD |
| Anna Karenina | *α*=0.5 | 0.57531 | 0.24639 | 0.42630 | 0.63870 | 0.24100 | **0.69271** |
| | *α*=1 | 0.48357 | 0.11719 | 0.27480 | 0.41424 | 0.11373 | **0.52205** |
| | *α*=1.5 | 0.43877 | 0.08775 | 0.22382 | 0.33808 | 0.08498 | **0.45620** |
| David Copperfield | *α*=0.5 | 0.55557 | 0.20533 | 0.38847 | 0.53029 | 0.19942 | **0.65528** |
| | *α*=1 | 0.46445 | 0.09499 | 0.26341 | 0.31133 | 0.09151 | **0.49349** |
| | *α*=1.5 | 0.42032 | 0.07067 | 0.21835 | 0.24619 | 0.06796 | **0.43238** |
| Political Books | *α*=0.5 | 0.63534 | 0.61968 | 0.63254 | 0.74574 | 0.60111 | **0.78163** |
| | *α*=1 | 0.58100 | 0.40081 | 0.46424 | 0.53996 | 0.38381 | **0.64311** |
| | *α*=1.5 | 0.55082 | 0.32719 | 0.39660 | 0.45883 | 0.31189 | **0.58493** |
| Book-Crossing | *α*=0.5 | 0.24497 | -0.00262 | 0.14004 | **0.31513** | -0.00334 | 0.30291 |
| | *α*=1 | 0.11798 | -0.00100 | 0.06207 | **0.15544** | -0.00131 | 0.15276 |
| | *α*=1.5 | 0.08856 | -0.00071 | 0.04574 | **0.11733** | -0.00095 | 0.11598 |
| CIAO | *α*=0.5 | 0.81492 | 0.62674 | 0.72950 | 0.81584 | 0.61389 | **0.82544** |
| | *α*=1 | 0.68009 | 0.41535 | 0.60891 | 0.63926 | 0.40626 | **0.68893** |
| | *α*=1.5 | 0.61488 | 0.34189 | 0.55058 | 0.56137 | 0.33478 | **0.62381** |
| Epinions | *α*=0.5 | 0.90853 | 0.85714 | 0.81085 | **0.95398** | 0.85754 | 0.93137 |
| | *α*=1 | 0.90685 | 0.75421 | 0.74352 | 0.89238 | 0.75788 | **0.92225** |
| | *α*=1.5 | 0.90578 | 0.70035 | 0.70594 | 0.85691 | 0.70542 | **0.91888** |
| MovieLens Latest | *α*=0.5 | 0.69222 | 0.10750 | 0.58095 | 0.54343 | 0.14731 | **0.70809** |
| | *α*=1 | 0.54531 | 0.04706 | 0.44395 | 0.32254 | 0.06787 | **0.54910** |
| | *α*=1.5 | 0.48001 | 0.03460 | 0.38565 | 0.25587 | 0.05050 | **0.48231** |
| Movie-Tweetings | *α*=0.5 | 0.87517 | 0.70382 | 0.80678 | **0.90473** | 0.69011 | 0.89339 |
| | *α*=1 | 0.79242 | 0.49575 | 0.66980 | 0.79160 | 0.47950 | **0.80843** |
| | *α*=1.5 | 0.74716 | 0.41684 | 0.60405 | 0.73286 | 0.40118 | **0.76307** |
| Football | *α*=0.5 | 0.60112 | 0.51887 | 0.75112 | 0.77096 | 0.27844 | **0.84888** |
| | *α*=1 | 0.57298 | 0.30899 | 0.59471 | 0.57390 | 0.13882 | **0.73124** |
| | *α*=1.5 | 0.55644 | 0.24550 | 0.52467 | 0.49312 | 0.10514 | **0.67374** |
| Primary School – Day1 | *α*=0.5 | 0.54967 | 0.14951 | 0.63216 | 0.54890 | 0.08783 | **0.78216** |
| | *α*=1 | 0.58420 | 0.07225 | 0.51312 | 0.32744 | 0.03743 | **0.71175** |
| | *α*=1.5 | 0.60905 | 0.05462 | 0.45785 | 0.26017 | 0.02737 | **0.68485** |
| Primary School – Day2 | *α*=0.5 | 0.60242 | 0.17388 | 0.66377 | 0.57006 | 0.12019 | **0.77386** |
| | *α*=1 | 0.61849 | 0.09768 | 0.55885 | 0.34661 | 0.05296 | **0.70531** |
| | *α*=1.5 | 0.62939 | 0.07838 | 0.50743 | 0.27701 | 0.03900 | **0.67779** |
| UK-Faculty | *α*=0.5 | 0.75712 | 0.55102 | 0.48789 | 0.65751 | 0.53896 | **0.79042** |
| | *α*=1 | 0.58760 | 0.33183 | 0.28865 | 0.43450 | 0.32077 | **0.60946** |
| | *α*=1.5 | 0.51388 | 0.26455 | 0.22876 | 0.35693 | 0.25475 | **0.53293** |
| Workplace | *α*=0.5 | 0.66406 | 0.31309 | 0.53201 | 0.59310 | 0.25743 | **0.74052** |
| | *α*=1 | 0.50255 | 0.15761 | 0.32668 | 0.36871 | 0.12690 | **0.54481** |
| | *α*=1.5 | 0.43480 | 0.11965 | 0.26189 | 0.29679 | 0.09587 | **0.46842** |
| AdjNoun | *α*=0.5 | 0.44578 | 0.14756 | 0.38114 | 0.32212 | 0.05982 | **0.51988** |
| | *α*=1 | 0.33611 | 0.06541 | 0.24235 | 0.15988 | 0.02498 | **0.35636** |
| | *α*=1.5 | 0.29037 | 0.04821 | 0.19649 | 0.12087 | 0.01819 | **0.29976** |
| +/=/- | *α*=0.5 | 14/0/0 | 14/0/0 | 14/0/0 | 11/0/3 | 14/0/0 | - |
| | *α*=1 | 14/0/0 | 14/0/0 | 14/0/0 | 13/0/1 | 14/0/0 | - |
| | *α*=1.5 | 14/0/0 | 14/0/0 | 14/0/0 | 13/0/1 | 14/0/0 | - |

**Table 5** The mean and the final ranks of the 6 algorithms of the second experiment according to the Friedman test, with considering the different values of *α*

| Algorithms | *α*=0.5 | | *α*=1 | | *α*=1.5 | |
|---|---|---|---|---|---|---|
| | Mean Rank | Final Rank | Mean Rank | Final Rank | Mean Rank | Final Rank |
| EM-BBO | 4.29 | 3 | 4.79 | 2 | 4.93 | 2 |
| SimAtt-BBO | 2.00 | 5 | 2.00 | 5 | 1.93 | 5 |
| CESNA | 3.21 | 4 | 3.14 | 4 | 3.21 | 4 |
| Topic-oriented | 4.43 | 2 | 3.86 | 3 | 3.79 | 3 |
| OV-SimAtt-BBO | 1.29 | 6 | 1.29 | 6 | 1.21 | 6 |
| MOBBO-OCD | 5.79 | 1 | 5.93 | 1 | 5.93 | 1 |

We subsequently performed a post-hoc test using the Wilcoxon signed-rank test implemented in IBM SPSS Statistics 22 at the significance level of 0.05, with considering the different values of *α*, to examine where the differences between the performances of the algorithms, that the Friedman test indicates, actually occur. With considering the different values of $\alpha$, Table 6 shows the $p$-values of the 6 algorithms of the second experiment against each other. In Table 6, the $p$-values below the Bonferroni-corrected critical value, which indicate the significant differences between the performances of the corresponding algorithms, are highlighted in bold-face. According to Table 6, compared with the other 5 algorithms, MOBBO-OCD obtains the largest number of $p$-values lower than the significance level with considering the Bonferroni correction, for all different values of $\alpha$. This indicates that the performance of MOBBO-OCD is significantly different from those of the most of the other algorithms of the experiment.

**Table 6** The $p$-values of the Wilcoxon signed-rank test of the 6 algorithms of the second experiment against each other at the significance level of 0.05, with considering the different values of $\alpha$. The $p$-values below the Bonferroni-corrected critical value are highlighted in bold-face.

| | | EM-BBO | SimAtt-BBO | CESNA | Topic-oriented | OV-SimAtt-BBO |
|---|---|---|---|---|---|---|
| **$\alpha$=0.5** | SimAtt-BBO | **9.82E-04** | - | - | - | - |
| | CESNA | 3.55E-02 | **2.87E-03** | - | - | - |
| | Topic-oriented | 9.75E-01 | **9.82E-04** | 2.58E-02 | - | - |
| | OV-SimAtt-BBO | **9.82E-04** | 7.63E-03 | **2.87E-03** | **9.82E-04** | - |
| | MOBBO-OCD | **9.82E-04** | **9.82E-04** | **9.82E-04** | 6.32E-03 | **9.82E-04** |
| **$\alpha$=1** | SimAtt-BBO | **9.82E-04** | - | - | - | - |
| | CESNA | **1.23E-03** | **1.89E-03** | - | - | - |
| | Topic-oriented | 3.51E-03 | **9.82E-04** | 5.10E-01 | - | - |
| | OV-SimAtt-BBO | **9.82E-04** | 1.32E-02 | **1.89E-03** | **9.82E-04** | - |
| | MOBBO-OCD | **9.82E-04** | **9.82E-04** | **9.82E-04** | **1.23E-03** | **9.82E-04** |
| **$\alpha$=1.5** | SimAtt-BBO | **9.82E-04** | - | - | - | - |
| | CESNA | **9.82E-04** | **1.52E-03** | - | - | - |
| | Topic-oriented | **1.52E-03** | **9.82E-04** | 7.78E-01 | - | - |
| | OV-SimAtt-BBO | **9.82E-04** | 1.32E-02 | **1.52E-03** | **9.82E-04** | - |
| | MOBBO-OCD | **9.82E-04** | **9.82E-04** | **9.82E-04** | **1.23E-03** | **9.82E-04** |

### 5.6. Third experiment

In the third experiment of this research, the performance of MOBBO-OCD is compared with those of the 2 recent community detection algorithms, i.e., semantic network-based community detection algorithm [30, 38] and SNTOCD (Semantic Network-based Topical Overlapping Community Detection) [30]. These 2 algorithms consider both the graph structure and the contents of a rating-based social network in their processes of community detection. As previously mentioned, semantic network-based community detection algorithm is an adaptation of the previously introduced algorithm of [38] for rating-based social networks, while SNTOCD is a general overlapping community detection framework, introduced in [30], with special focus on rating-based social networks. Thus, in this experiment, the performance of MOBBO-OCD is compared with those of the 2 competitor algorithms of this experiment on the 5 rating-based data sets described in Sub-section 5.1, which are Book-Crossing, CIAO, Epinions, Movielens Latest and Movie-Tweetings.

Like the previous experiments, for each of the 2 competitor algorithms of MOBBO-OCD in this experiment, we used the codes, software or information provided by its corresponding researchers. Since *alpha_SAEM* can evaluate the performance of community detection in undirected unweighted social networks, and is not able to handle weighted and directed networks, the achieved edge weights in the process of semantic network-based community detection algorithm and SNTOCD, which are more than 1, are set to 1.

Table 7 presents the mean of the best-of-run *alpha_SAEM* values, for $\alpha$=0.5, $\alpha$=1 and $\alpha$=1.5, over independent runs of MOBBO-OCD and the other 2 algorithms of the third experiment on the 5 rating-based data sets of the experiments. According to the last row of Table 7, MOBBO-OCD notably outperforms all of its competitors in the third experiment. According to Table 7, for all values of $\alpha$, MOBBO-OCD shows the absolute best performance among its 2 competitors on all the 5 rating-based data sets of the experiments. This indicates that the performance of MOBBO-OCD is quite superior to the performances of the other algorithms in the third experiment.

**Table 7** The mean of the best-of-run *alpha_SAEM* values, for $\alpha$=0.5, $\alpha$=1 and $\alpha$ =1.5, over independent runs of MOBBO-OCD and the other 2 algorithms of the third experiment on the 5 rating-based data sets of the experiments

| Data Sets | alpha_SAEM | Algorithms | | |
|---|---|---|---|---|
| | | Semantic network-based | SNTOCD | MOBBO-OCD |
| Book-Crossing | $\alpha$=0.5 | 0.18264 | 0.17040 | **0.30291** |
| | $\alpha$=1 | 0.08385 | 0.07594 | **0.15276** |

| | $\alpha$=1.5 | 0.06227 | 0.05603 | **0.11598** |
|---|---|---|---|---|
| CIAO | $\alpha$=0.5 | 0.73850 | 0.72731 | **0.82544** |
| | $\alpha$=1 | 0.56206 | 0.51617 | **0.68893** |
| | $\alpha$=1.5 | 0.48741 | 0.43519 | **0.62381** |
| Epinions | $\alpha$=0.5 | 0.84162 | 0.84995 | **0.93137** |
| | $\alpha$=1 | 0.74596 | 0.69380 | **0.92225** |
| | $\alpha$=1.5 | 0.69529 | 0.62071 | **0.91888** |
| MovieLens Latest | $\alpha$=0.5 | 0.47043 | 0.33443 | **0.70809** |
| | $\alpha$=1 | 0.31164 | 0.16735 | **0.54910** |
| | $\alpha$=1.5 | 0.25621 | 0.12676 | **0.48231** |
| Movie-Tweetings | $\alpha$=0.5 | 0.79082 | 0.80913 | **0.89339** |
| | $\alpha$=1 | 0.65512 | 0.62904 | **0.80843** |
| | $\alpha$=1.5 | 0.59021 | 0.55050 | **0.76307** |
| +/=/- | $\alpha$=0.5 | 5/0/0 | 5/0/0 | - |
| | $\alpha$=1 | 5/0/0 | 5/0/0 | - |
| | $\alpha$=1.5 | 5/0/0 | 5/0/0 | - |

The results of the third experiment, shown in Table 7, and the above analysis of the results can demonstrate the best performance of MOBBO-OCD among its 2 competitors in the experiment, but to ensure that the performance of MOBBO-OCD is statistically different from those of the other 2 algorithms, we conducted a statistical significance analysis on the results in Table 7, with considering the different values of $\alpha$, using the Friedman test implemented in IBM SPSS Statistics 22. The analysis shows that there is a statistically significant difference between the performances of the algorithms with $\chi^2(2) = 7.600000$ and $p < 0.05$ for $\alpha = 0.5$, $\chi^2(2) = 10.000000$ and $p < 0.05$ for $\alpha = 1$, $\chi^2(2) = 10.000000$ and $p < 0.05$ for $\alpha = 1.5$. Also, the result of Friedman ranking, which is shown in Table 8, indicates that MOBBO-OCD ranks first among all the algorithms in the experiment, with the mean rank value of 3.00 for all values of $\alpha$.

**Table 8** The mean and the final ranks of the 3 algorithms of the third experiment according to the Friedman test, with considering the different values of $\alpha$

| Algorithms | $\alpha$=0.5 | | $\alpha$=1 | | $\alpha$=1.5 | |
|---|---|---|---|---|---|---|
| | Mean Rank | Final Rank | Mean Rank | Final Rank | Mean Rank | Final Rank |
| Semantic network-based | 1.60 | 2 | 2.00 | 2 | 2.00 | 2 |
| SNTOCD | 1.40 | 3 | 1.00 | 3 | 1.00 | 3 |
| MOBBO-OCD | 3.00 | 1 | 3.00 | 1 | 3.00 | 1 |

We subsequently performed a post-hoc test using the Wilcoxon signed-rank test implemented in IBM SPSS Statistics 22 at the significance level of 0.05, with considering the different values of $\alpha$. Table 9 shows the $p$-values of the 3 algorithms of the third experiment against each other, with considering the different values of $\alpha$. In Table 9, the $p$-values below the significance level are highlighted in bold-face. According to Table 9, compared with the other 2 algorithms, MOBBO-OCD obtains the largest number of $p$-values lower than the significance level, for all different values of $\alpha$.

**Table 9** The $p$-values of the Wilcoxon signed-rank test of the 3 algorithms of the third experiment against each other at the significance level of 0.05, with considering the different values of $\alpha$. The $p$-values below the significance level are highlighted in bold-face.

| | | Semantic network-based | SNTOCD |
|---|---|---|---|
| **$\alpha$=0.5** | SNTOCD | 5.00E-01 | - |
| | MOBBO-OCD | **4.31E-02** | **4.31E-02** |
| **$\alpha$=1** | SNTOCD | **4.31E-02** | - |
| | MOBBO-OCD | **4.31E-02** | **4.31E-02** |
| **$\alpha$=1.5** | SNTOCD | **4.31E-02** | - |
| | MOBBO-OCD | **4.31E-02** | **4.31E-02** |

### 5.7. Summarization of the analyses

In this sub-section, we present two figures to summarize the analyses conducted in the three experiments of this research. Fig. 3 shows the performance comparison of the 14 algorithms of the first and the second experiments by averaging their achieved mean values of *alpha_SAEM* for $\alpha$=0.5, $\alpha$=1 and $\alpha$=1.5, which are presented in Table 1 and Table 4, on the 14 data sets of the experiments.

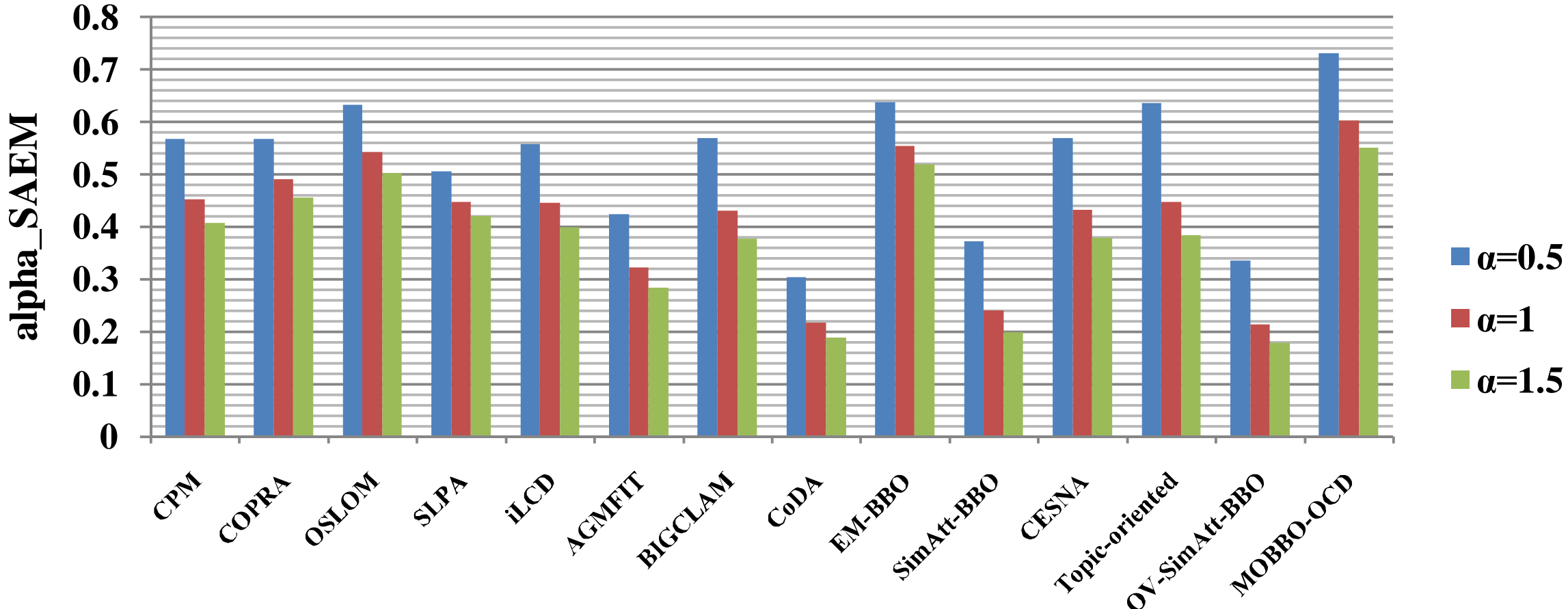

**Fig. 3.** The performance comparison of the 14 algorithms of the first and the second experiments by considering the averages of their achieved mean values of *alpha_SAEM*, for $\alpha$=0.5, $\alpha$=1 and $\alpha$=1.5, on the 14 data sets of the experiments

Also, Fig. 4 shows the performance comparison of the 3 algorithms of the third experiment by averaging their achieved mean values of *alpha_SAEM* for $\alpha$=0.5, $\alpha$=1 and $\alpha$=1.5, which are presented in Table 7, on the 5 rating-based data sets of the experiments.

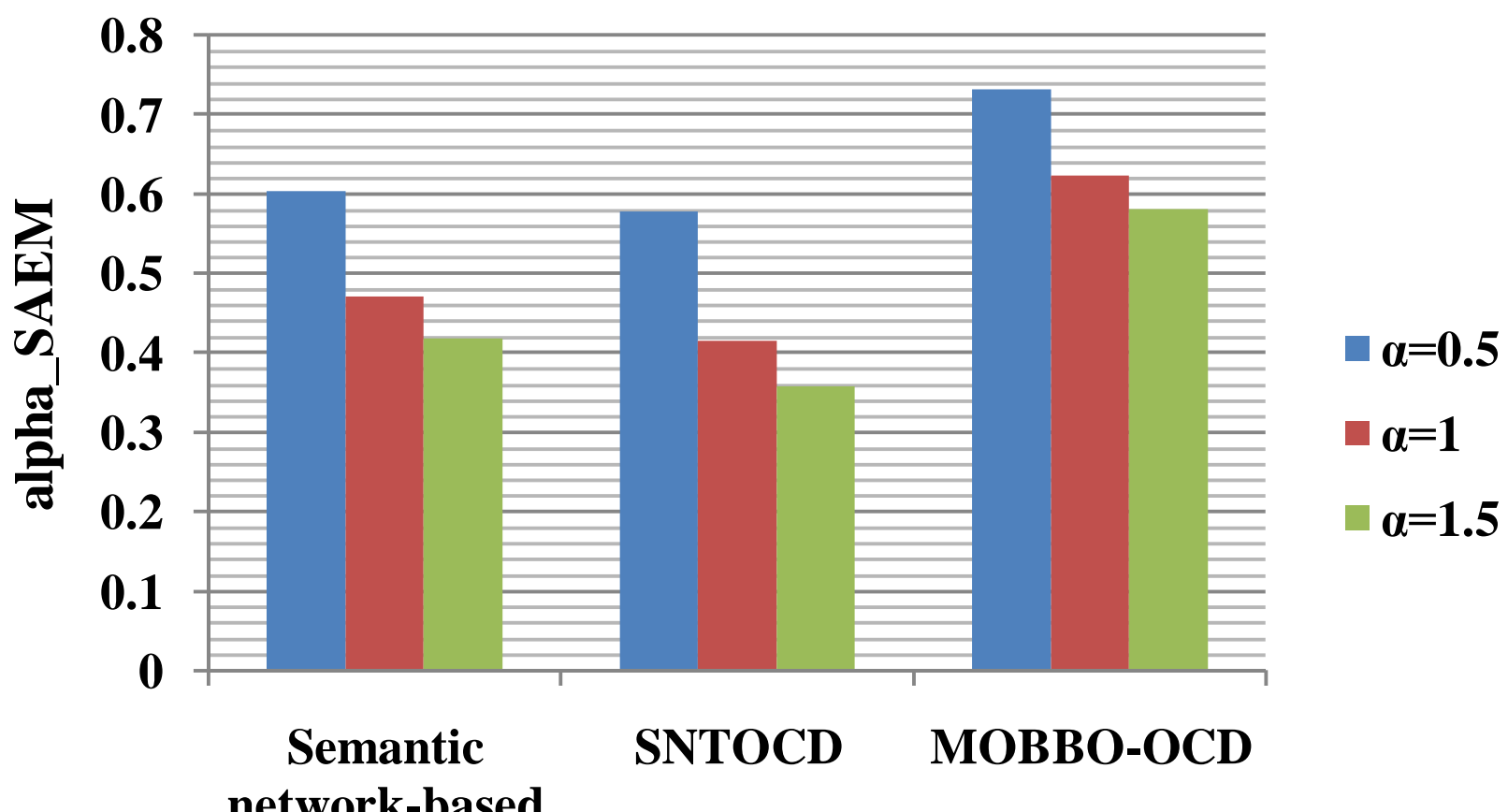

**Fig. 4.** The performance comparison of the 3 algorithms of the third experiment by considering the averages of their achieved mean values of *alpha_SAEM*, for $\alpha$=0.5, $\alpha$=1 and $\alpha$=1.5, on the 5 rating-based data sets of the experiments

According to Fig. 3 and Fig. 4, it can be concluded that MOBBO-OCD achieves the best average performance among its competitors in all of these three conditions: 1) When the similarity of nodes' attributes is as important for us as the link ( $\alpha = 1$ ), 2) When the similarity of nodes' attributes is more important for us than the link ( $0 \leq \alpha < 1$ ), and 3) When the link is more important for us than the similarity of nodes' attributes ( $\alpha > 1$ ).

## 6. Conclusion

In this paper, a Multi-Objective BBO-based Overlapping Community Detection algorithm called MOBBO-OCD has been proposed for automatic detection of overlapping communities of a social network, in which node attributes are available, with considering the two aspects of topological structure and node attributes of the network. *SimAtt*, which considers the similarity of nodes' attributes, and Extended Modularity, which considers the density of connections, have been chosen as the two objective functions to be maximized in MOBBO-OCD. Since MOBBO-OCD uses the Pareto-based approach, its final output is a set of non-dominated solutions (partitions) of its input social network. Thus, our proposed method can provide a wide range of solutions for a decision maker to choose from. For this reason, a metric called *alpha_SAEM* has been introduced in this paper to determine the best compromise solution among the set of non-dominated solutions achieved by a community detection algorithm. *alpha_SAEM* is able to evaluate the goodness of both overlapping and non-overlapping partitions with considering the two aspects of node attributes and linkage structure.

In this paper, three extensive experiments have been conducted on 14 real-life data sets with different characteristics to evaluate the performance of MOBBO-OCD by comparing its results with those of 15 relevant community detection algorithms. In the first experiment, the performance of MOBBO-OCD has been compared with those of the 8 state-of-the-art overlapping community detection algorithms, which conduct their community detection processes with just considering the graph structures of networks. In the second experiment, the performance of MOBBO-OCD has been compared with Extended Modularity maximization BBO algorithm (EM-BBO) along with the 4 overlapping community detection algorithms, which consider both the graph structure and the contents of a network in their processes of community detection. In the third experiment, the performance of MOBBO-OCD has been compared with those of the 2 recent community detection algorithms, which consider both the graph structure and the contents of a rating-based social network in their processes of community detection. The experimental results, which have been statistically validated, show that MOBBO-OCD achieves favorable results which are quite superior to the results of the other algorithms in the experiments. Thus, it can be concluded that MOBBO-OCD can be considered as an efficient method to find overlapping communities in social networks with node attributes with synchronously considering structure and attribute.

Like many other proposed frameworks, MOBBO-OCD, has some limitations that can be considered as potential research directions for future studies. MOBBO-OCD, in its current form, can only be applied to unweighted undirected social networks with node attributes. But the proposed framework can be considered as an adaptive general framework in the field of community detection. Thus, In future works, MOBBO-OCD can be extended to detect overlapping communities of weighted and directed social networks with node attributes. Since discovering communities of a signed network is a new interesting research area in the field of community detection, another promising avenue for future studies can be the enhancement of MOBBO-OCD to detect communities of signed networks with node attributes. Another interesting research area in the field of community detection is the analysis of the evolution of communities over time. Thus, another direction for future works can be the enhancement of MOBBO-OCD to analyze the trend of evolution in social networks with node attributes, over time. On the other hand, the concepts and approaches, which have been discussed in this study, can be utilized in future studies, to design new frameworks for link prediction in social networks with node attributes.

## References

[1] Friedman Test in SPSS Statistics - How to run the procedure, understand the output using a relevant example | Laerd statistics., in: https://statistics.laerd.com/spss-tutorials/friedman-test-using-spss-statistics.php.

[2] B. Amiri, L. Hossain, J.W. Crawford, R.T. Wigand, Community detection in complex networks: Multi–objective enhanced firefly algorithm, Knowledge-Based Systems, 46 (2013) 1-11.

[3] H. Bostani, M. Sheikhan, Modification of supervised OPF-based intrusion detection systems using unsupervised learning and social network concept, Pattern Recognition, 62 (2017) 56-72.

[4] R. Cazabet, F. Amblard, Simulate to detect: a multi-agent system for community detection, in: Proceedings of the 2011 IEEE/WIC/ACM International Conferences on Web Intelligence and Intelligent Agent Technology-Volume 02, IEEE Computer Society, 2011, pp. 402-408.

[5] D. Chen, F. Zou, R. Lu, L. Yu, Z. Li, J. Wang, Multi-objective optimization of community detection using discrete teaching–learning-based optimization with decomposition, Information Sciences, (2016).

[6] H. Chen, Z. Yu, Q. Yang, J. Shao, Community detection in subspace of attribute, Information Sciences, 602 (2022) 220-235.

[7] C.A.C. Coello, D.A. Van Veldhuizen, G.B. Lamont, Evolutionary algorithms for solving multi-objective problems, Springer, 2002.

[8] K. Deb, A. Pratap, S. Agarwal, T. Meyarivan, A fast and elitist multiobjective genetic algorithm: NSGA-II, IEEE transactions on evolutionary computation, 6 (2002) 182-197.

[9] X. Ding, H. Yang, J. Zhang, J. Yang, X. Xiang, CEO: Identifying Overlapping Communities via Construction, Expansion and Optimization, Information Sciences, 596 (2022) 93-118.

[10] O. Doluca, K. Oğuz, APAL: Adjacency Propagation Algorithm for overlapping community detection in biological networks, Information Sciences, 579 (2021) 574-590.

[11] S. Gregory, Finding overlapping communities in networks by label propagation, New Journal of Physics, 12 (2010) 103018.

[12] J. Handl, J. Knowles, An evolutionary approach to multiobjective clustering, IEEE transactions on Evolutionary Computation, 11 (2007) 56-76.

[13] C. He, Y. Zheng, J. Cheng, Y. Tang, G. Chen, H. Liu, Semi-supervised overlapping community detection in attributed graph with graph convolutional autoencoder, Information Sciences, 608 (2022) 1464-1479.

[14] E. Jokar, M. Mosleh, Community detection in social networks based on improved label propagation algorithm and balanced link density, Physics Letters A, 383 (2019) 718-727.

[15] E. Jokar, M. Mosleh, M. Kheyrandish, Overlapping community detection in complex networks using fuzzy theory, balanced link density, and label propagation, Expert Systems, (2021) e12921.

[16] A. Lancichinetti, S. Fortunato, Consensus clustering in complex networks, Scientific reports, 2 (2012).

[17] A. Lancichinetti, F. Radicchi, J.J. Ramasco, S. Fortunato, Finding statistically significant communities in networks, PloS one, 6 (2011) e18961.

[18] Z. Li, J. Liu, K. Wu, A multiobjective evolutionary algorithm based on structural and attribute similarities for community detection in attributed networks, IEEE transactions on cybernetics, 48 (2018) 1963-1976.

[19] D. Liu, Z. Chang, G. Yang, E. Chen, Hiding ourselves from community detection through genetic algorithms, Information Sciences, 614 (2022) 123-137.

[20] H. Ma, An analysis of the equilibrium of migration models for biogeography-based optimization, Information Sciences, 180 (2010) 3444-3464.

[21] H. Ma, Z. Liu, X. Zhang, L. Zhang, H. Jiang, Balancing topology structure and node attribute in evolutionary multi-objective community detection for attributed networks, Knowledge-Based Systems, 227 (2021) 107169.

[22] H. Ma, D. Simon, P. Siarry, Z. Yang, M. Fei, Biogeography-based optimization: a 10-year review, IEEE Transactions on Emerging Topics in Computational Intelligence, 1 (2017) 391-407.

[23] H. Ma, H. Yang, K. Zhou, L. Zhang, X. Zhang, A local-to-global scheme-based multi-objective evolutionary algorithm for overlapping community detection on large-scale complex networks, Neural Computing and Applications, 33 (2021) 5135-5149.

[24] M.E. Newman, M. Girvan, Finding and evaluating community structure in networks, Physical review E, 69 (2004) 026113.

[25] G. Palla, I. Derényi, I. Farkas, T. Vicsek, Uncovering the overlapping community structure of complex networks in nature and society, Nature, 435 (2005) 814-818.

[26] A. Papadopoulos, G. Pallis, M.D. Dikaiakos, Weighted clustering of attributed multi-graphs, Computing, 99 (2017) 813-840.

[27] M. Qin, K. Lei, Dual-channel hybrid community detection in attributed networks, Information Sciences, 551 (2021) 146-167.

[28] U.N. Raghavan, R. Albert, S. Kumara, Near linear time algorithm to detect community structures in large-scale networks, Physical review E, 76 (2007) 036106.

[29] A. Reihanian, M.-R. Feizi-Derakhshi, H.S. Aghdasi, Community detection in social networks with node attributes based on multi-objective biogeography based optimization, Engineering Applications of Artificial Intelligence, 62 (2017) 51-67.

[30] A. Reihanian, M.-R. Feizi-Derakhshi, H.S. Aghdasi, Overlapping community detection in rating-based social networks through analyzing topics, ratings and links, Pattern Recognition, 81 (2018) 370-387.

[31] A. Reihanian, B. Minaei-Bidgoli, H. Alizadeh, Topic-oriented community detection of rating-based social networks, Journal of King Saud University-Computer and Information Sciences, 28 (2016) 303-310.

[32] L.C. Ribas, J.J.d.M.S. Junior, L.F. Scabini, O.M. Bruno, Fusion of complex networks and randomized neural networks for texture analysis, Pattern Recognition, 103 (2020) 107189.

[33] R. Shang, K. Zhao, W. Zhang, J. Feng, Y. Li, L. Jiao, Evolutionary multiobjective overlapping community detection based on similarity matrix and node correction, Applied Soft Computing, 127 (2022) 109397.

[34] H. Shen, X. Cheng, K. Cai, M.-B. Hu, Detect overlapping and hierarchical community structure in networks, Physica A: Statistical Mechanics and its Applications, 388 (2009) 1706-1712.

[35] D. Simon, Biogeography-based optimization, IEEE transactions on evolutionary computation, 12 (2008) 702-713.

[36] D. Simon, R. Rarick, M. Ergezer, D. Du, Analytical and numerical comparisons of biogeography-based optimization and genetic algorithms, Information Sciences, 181 (2011) 1224-1248.

[37] Y. Tian, S. Yang, X. Zhang, An evolutionary multiobjective optimization based fuzzy method for overlapping community detection, IEEE Transactions on Fuzzy Systems, 28 (2019) 2841-2855.

[38] Z. Xia, Z. Bu, Community detection based on a semantic network, Knowledge-Based Systems, 26 (2012) 30-39.

[39] J. Xie, B.K. Szymanski, X. Liu, Slpa: Uncovering overlapping communities in social networks via a speaker-listener interaction dynamic process, in: Data Mining Workshops (ICDMW), 2011 IEEE 11th International Conference on, IEEE, 2011, pp. 344-349.

[40] Y. Xing, F. Meng, Y. Zhou, G. Sun, Z. Wang, Overlapping Community Detection Extended from Disjoint Community Structure, Computing and Informatics, 38 (2019) 1091-1110.

[41] J. Yang, J. Leskovec, Community-affiliation graph model for overlapping network community detection, in: 2012 IEEE 12th International Conference on Data Mining, IEEE, 2012, pp. 1170-1175.

[42] J. Yang, J. Leskovec, Overlapping community detection at scale: a nonnegative matrix factorization approach, in: Proceedings of the sixth ACM international conference on Web search and data mining, ACM, 2013, pp. 587-596.

[43] J. Yang, J. McAuley, J. Leskovec, Community detection in networks with node attributes, in: 2013 IEEE 13th International Conference on Data Mining, IEEE, 2013, pp. 1151-1156.

[44] J. Yang, J. McAuley, J. Leskovec, Detecting cohesive and 2-mode communities indirected and undirected networks, in: Proceedings of the 7th ACM international conference on Web search and data mining, ACM, 2014, pp. 323-332.

[45] Y. Yang, P. Shi, Y. Wang, K. He, Quadratic Optimization based Clique Expansion for overlapping community detection, Knowledge-Based Systems, 247 (2022) 108760.

[46] L. Zhang, H. Pan, Y. Su, X. Zhang, Y. Niu, A mixed representation-based multiobjective evolutionary algorithm for overlapping community detection, IEEE Transactions on Cybernetics, 47 (2017) 2703-2716.

[47] Z. Zhao, S. Feng, Q. Wang, J.Z. Huang, G.J. Williams, J. Fan, Topic oriented community detection through social objects and link analysis in social networks, Knowledge-Based Systems, 26 (2012) 164-173.

[48] Z. Zhao, Z. Ke, Z. Gou, H. Guo, K. Jiang, R. Zhang, The trade-off between topology and content in community detection: An adaptive encoder–decoder-based NMF approach, Expert Systems with Applications, 209 (2022) 118230.

[49] X. Zhou, Y. Liu, B. Li, G. Sun, Multiobjective biogeography based optimization algorithm with decomposition for community detection in dynamic networks, Physica A: Statistical Mechanics and its Applications, 436 (2015) 430-442.

[50] X. Zhou, L. Su, X. Li, Z. Zhao, C. Li, Community detection based on unsupervised attributed network embedding, Expert Systems with Applications, 213 (2023) 118937.